\documentclass[fleqn,usenatbib]{mnras}

\usepackage{newtxtext,newtxmath}

\usepackage[T1]{fontenc}
\usepackage{ae,aecompl}

\usepackage{graphicx}	
\usepackage{amsmath}	
\usepackage{amssymb}	
\usepackage{gensymb}
\usepackage{textcomp}	
\usepackage{pdflscape}
\usepackage{enumitem}

\defcitealias{nav16}{Paper I}
\defcitealias{nish09}{NTH09}
\defcitealias{woo98}{WHM98}
\defcitealias{mkn09}{MKN09}
\defcitealias{mfk13}{MFK13}
\defcitealias{nish06}{NNK06}
\defcitealias{alon17}{AMC17}
\newcommand{\msun}{M_{\odot}}

\newcommand{\rsun}{R_{\odot}}

\title[Long-term variability in the Galactic centre]{Long-term Stellar Variability in the Galactic Centre Region}

\author[C. Navarro Molina et al.]{
Claudio Navarro Molina,$^{1}$\thanks{E-mail: claudio.navarro@postgrado.uv.cl (CNM)}
J. Borissova,$^{2,1}$
M. Catelan,$^{3,1}$\thanks{On sabbatical leave at at the Astronomisches Rechen-Institut, Zentrum f\"ur Astronomie der Universit\"at Heidelberg, M\"onchofstrasse 12-14, 69120 Heidelberg, Germany.}
P. W. Lucas,$^{4}$
\newauthor N. Medina$^{1},$
C. Contreras Pe\~na,$^{5}$
R. Kurtev,$^{2,1}$
and D. Minniti,$^{1,6,7}$
\\
$^{1}$Millennium Institute of Astrophysics, Av. Vicu\~na Mackenna 4860, 782-0436 Macul, Santiago, Chile\\
$^{2}$Instituto de F\'{\i}sica y Astronom\'{\i}a, Universidad de Valpara\'{\i}so, Gran Breta\~na 1111, Casilla 5030, Valpara\'{\i}so, Chile\\
$^{3}$Instituto de Astrof\'{\i}sica, Facultad de F\'{\i}sica, Pontificia Universidad Cat\'olica de Chile, Casilla 306, Santiago 22, Chile\\
$^{4}$Centre for Astrophysics Research, Science and Technology Research Institute, University of Hertfordshire, Hatfield, AL10 9AB, UK\\
$^{5}$Department of Physics and Astronomy, University of Exeter, Stocker Road, Exeter, Devon EX4 4SB, UK\\
$^{6}$Departamento de Ciencias F\'{\i}sicas, Universidad Andr\'es Bello, Rep\'ublica 220, Santiago, Chile\\
$^{7}$Vatican Observatory, V00120 Vatican City State, Italy\\
}

\date{Accepted XXX. Received YYY; in original form ZZZ}

\pubyear{2018}

\begin{document}
\label{firstpage}
\pagerange{\pageref{firstpage}--\pageref{lastpage}}
\maketitle

\begin{abstract}
We report the detection of variable stars within a $11.5'\times11.5'$ region near the Galactic centre (GC) that includes the Arches and Quintuplet clusters, as revealed by the VISTA Variables in the V\'{i}a L\'actea (VVV) survey. There are 353 sources that show $K_S$-band variability, of which the large majority ($81\%$) correspond to red giant stars, mostly in the asymptotic giant branch (AGB) phase. We analyze a population of 52 red giants with long-term trends that cannot be classified into the typical pulsating star categories. Distances and extinctions are calculated for 9 Mira variables, and we discuss the impact of the chosen extinction law on the derived distances. We also report the presence of 48 new identified young stellar object (YSO) candidates in the region.
\end{abstract}

\begin{keywords}
stars: variables: general -- infrared: stars -- stars: AGB and post-AGB
\end{keywords}



\section{Introduction}

The Galactic bulge contains  a wide variety of stellar populations. While the external region, with a radius of 3 kpc, holds mainly stars older than $8-10$ Gyr \citep{zoc03,ebea18}, the central part, called the nuclear bulge \citep*{lau02}, contains both old and young stars. \citet{fig04} find a history of continuous star formation. This young population is mostly located inside three massive clusters located in the zone: Arches, Quintuplet and the Nuclear Cluster, all of them concentrated within 30 pc around the GC. The nuclear bulge extends up to $\sim 200$ pc, and is embedded in interstellar material, known as the central molecular zone \citep[CMZ;][]{mor96}.

The knowledge of the surrounding stellar population is crucial to understand how this extremely dense environment affects the formation and evolution of stars. Near-infrared surveys like the VVV \citep{min10, sai12} have helped to uncover the dust veil around this region affected by a very strong foreground extinction, which can be as high as $A_V \sim 40$ magnitudes \citep{nis08,fri11}.

Near-infrared variability is present in a wide range of stars, and its detection is one of the main VVV goals. \citet{cat13} show a summary of different variability classes found in this survey. Among these classes, one finds YSOs and AGB stars. YSOs are expected to show different levels of variability coming from a variety of physical processes, such as changes in the accretion rate, variable extinction, and hot or cool spots \citep[see, e.g.,][]{her94,cod14}. While the spots can lead to relatively small-amplitude variations, at the level of tenths of magnitudes in the $K$ band, abrupt changes in the accretion rate from the stellar disk can lead to an increase in brightness by $5-6$ magnitudes in the visual, as in the case of FU Orionis (FUor) stars. EX Lupi (EXor) stars also present abrupt variations in the accretion rate, but at a smaller scale. Lately, a new class, called MNor, has been suggested \citep{cont17b}, which shows mixed features from the two previous types. In general, YSO variability is rather irregular, with timescales ranging from a few days up to a decade, depending on the physical processes involved. 

On the other hand, AGB stars also show significant variability in near-infrared bands. Perhaps the best known among these variables are the Mira stars. They are low- and intermediate-mass stars that show fairly regular, almost sinusoidal light curves, with periods longer than 100 days and large visual amplitudes ($\Delta V > 2.5$~mag). Miras can be used as tracers of Galactic structure, thanks to their well-known period-luminosity (PL) relation \citep[e.g.,][]{gla81,whit08}. Depending on their surface C/O ratio, Miras separate into two groups. O-rich Miras contain molecules such as H$_2$O, TiO and SiO, while C-rich Miras contain C$_2$ and CN, with the C being dredged up from the stellar interior to the surface \citep{mats17}. In the Galactic bulge, the bulk of the Mira population is O-rich, while only a few C-rich Miras have been observed \citep{whit06,mats17}. In other low-metallicity environments, like the Magellanic Clouds, the fraction of C-rich Miras is larger \citep[e.g.,][]{whit03}.

Other variable types found in this evolutionary stage are semi-regular or irregular variables (hereafter SR and IRR, respectively), in addition to the so-called OGLE small-amplitude red giants (OSARGs). The former display a prominent periodicity in their light curve, but not as regular as for Miras, and the amplitudes are smaller as well; periods lie between 20 to 2000 days in most cases. Variables of the IRR type do not show clear signs of periodicity. Finally, the OSARGs \citep{wra04} are a variability type found by the Optical Gravitational Lensing Experiment \citep[OGLE;][]{auea92} survey. We must note that the distinction between these classes is not always clear, since proper classification hinges heavily on the availability of suitable, properly sampled time-series. For this reason, very often such variability classes are bundled into the ``long-period variables'' (LPV) name, even when strict periodicity cannot be established \citep[see, e.g.,][and references therein]{cat15}.

Even though Miras have been considered to have regular variations in time, secular changes in their periods and amplitudes have been found in some of them when they are observed for a very long time \citep[e.g.][]{smi13}. Sometimes, these changes are associated with thermal pulses during the AGB phase \citep*{tem05}. Additionally, \citet{zij02} noted that $\sim10\%$ of LPVs showed oscillating period variations on a timescale of decades. They called them ``meandering Miras''. \citet{tem05} relate these changes to thermal oscillations on a Kelvin-Helmholtz timescale.

More common than period changes are amplitude variations in LPVs. It has been reported that C-rich Miras show cycle-to-cycle variations more often than do O-rich Miras \citep[e.g.,][]{whit03,whit06,mats17}. \citet{whit03} detected a long-term trend in some light curves of C-rich Miras, though the periodicity of those trends could not be reliably estimated. For other C-rich Miras, there is evidence of dust obscuration events, like in the variable star R For \citep{whit97}. A possible explanation is that the creation of new dust in the circumstellar shell may form structures that would cause the deep minima observed in the light curve of R For \citep{wint94}. Also, \citet{whit97} make the analogy with R Coronae Borealis (RCB) stars, in the sense that these stars could be ejecting ``puffs'' of materials in random directions. Therefore, variable mass loss could account for the long-term variation of carbon Miras. Clearly, it is important to follow the temporal evolution of periodic and semi-periodic AGB stars in order to have a better understanding of this phenomenon.

Only a handful of variability surveys have focused on the GC. \citet*[][hereafter WHM98]{woo98} analyzed 109 OH/IR stars (O-rich Miras with large amplitudes and periods up to $\sim1000$ days), a sample mainly based on the VLA radio observations of \citet{lin92}. \citet{gla01} performed a 4-year $K$-band survey in a $24'\times 24'$ region around the GC; they found 409 LPVs. Later, \citet[][hereafter MKN09]{mkn09} surveyed a similar zone as \citeauthor{gla01}, this time in the $JHK_S$-bands. Thanks to deeper photometry, the number of variable stars found increased to 1364. Of these, distances and extinctions were calculated for 143 Miras, confirming that this population effectively belongs to the GC. But not only LPVs are present in this area.
\citet{mats11} discovered three classical Cepheids in the nuclear bulge, while a fourth one was added by \citet{mats15}. In addition, \citet[][hereafter MFK13]{mfk13} reported 45 short-period variables in the same region analyzed by \citetalias{mkn09}, most of them being type II Cepheids and eclipsing binaries (hereafter EBs). 

Here we present the results of a $K_S$-band variability search carried out with the VVV data in a region near the GC, in order to analyze the stellar population of this zone. 


\section{Data}
\label{sec:data}

\begin{figure*}
\centering
\includegraphics[width=\textwidth]{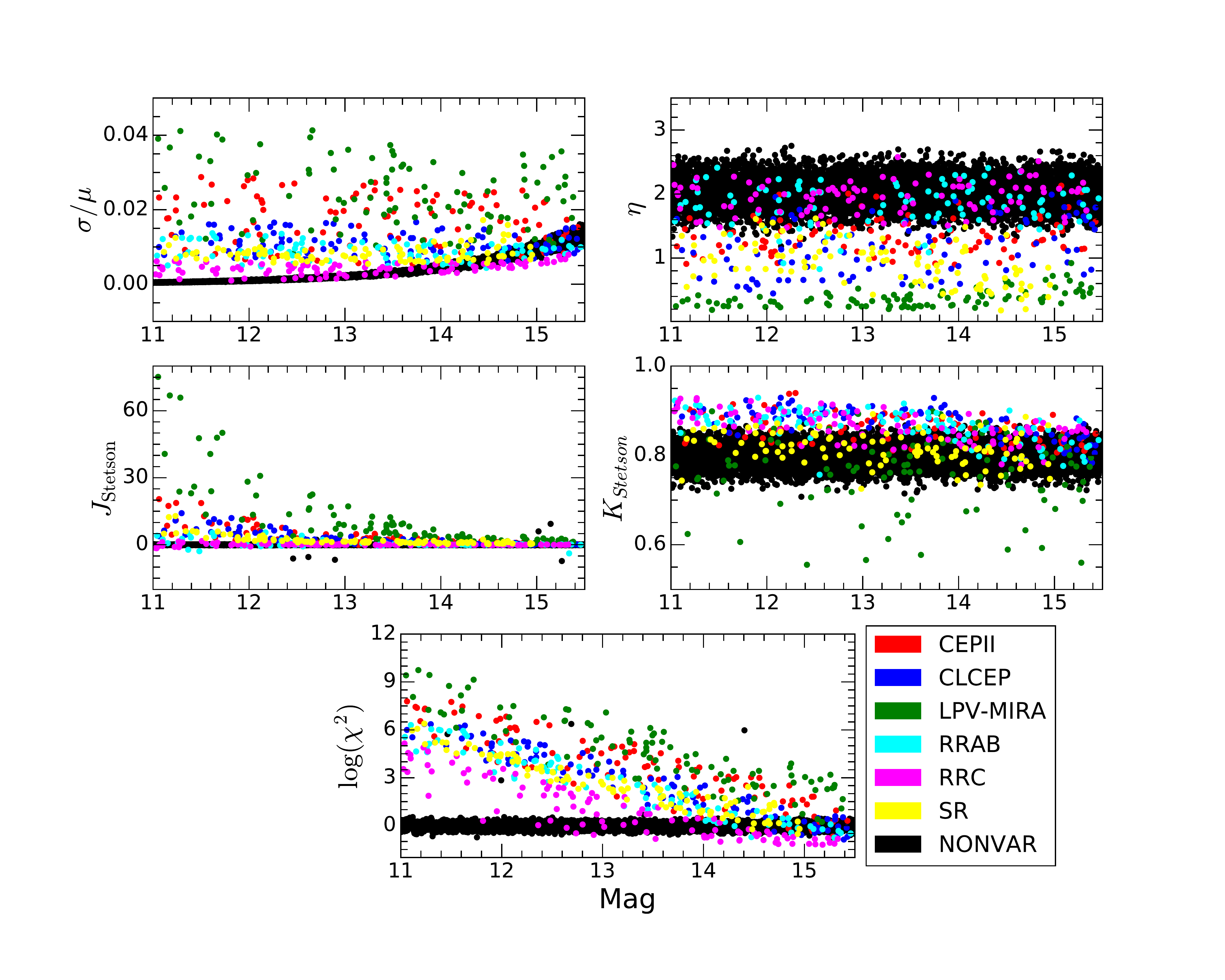}
\caption{Example of performance test of five variability indexes against artificial stars: mean variance ({\it top left}), von Neumann ratio ({\it top right}), Stetson $J$ ({\it middle left}) and $K$ ({\it middle right}), and the $\log {\chi^2}$ index ({\it bottom}). Each colour corresponds to a different variability type, namely type II Cepheids (red), classical Cepheids (blue), Miras (green), RRab (cyan), RRc (magenta), and SR variables (yellow). Black circles correspond to 10,000 non-variable stars. 
\label{fig-1}}
\end{figure*}

Data for this work were obtained from the VVV survey. Images were taken with the VIRCAM camera at the VISTA 4.1 m telescope, located in Cerro Paranal, Chile. This survey comprises $K_S$ images of the Galactic bulge and part of the disk, acquired between August 2010 and September 2015, in addition to two single and nearly simultaneous $ZYJH$ observations, taken during the first and last year of the project.

Details of the images used and the photometry performed are found in \citet[][hereafter Paper I]{nav16}. This time, a larger area ($11.5'\times11.5'$) was used, which included the Arches and Quintuplet star clusters. A total of 89 pawprint (i.e., a single exposure image) epochs are available, with a time range equal to the full survey duration. This  is a $\sim60\%$ increase in number of epochs compared to \citetalias{nav16}.

The final step after the photometric calibration was the construction of a master catalogue. All single epoch photometric catalogues were matched with an arbitrarily chosen reference epoch, which in this case corresponds to the image with the best seeing. Cross-match was performed using {\sc stilts}\footnote{Starlink Tables Infrastructure Library Tool Set}, with a tolerance set at 0.4\arcsec. Only objects with a {\sc dophot} flag of 1 (stellar object with good photometry) or 7 (faint object) were considered, thus excluding saturated, blended and non-stellar objects. The final master catalogue contains 97,659 stars.

\subsection{Artificial stars and templates}

To perform the search for variable stars, several different variability indexes were used. To analyze the effectivity of a given variability index we created 10,000 artificial light curves of non-variable stars, with magnitudes and errors similar to those found in our photometric catalogs. Besides, the epoch cadence was chosen to emulate the uneven temporal spacing of VVV observations. In addition, we generated light curves of some variable star classes. For this, data from the VVV Templates Project \citep{ange14} was employed. Again, magnitudes and errors were adapted to match typical VVV data. A total of 100 artificial variable stars were created per variability class: type II Cepheids, classical Cepheids, RR Lyrae stars of types RRab and RRc, Miras and SR variables. YSOs were not considered in this analysis due to the heterogeneous nature of their light curves, which is a consequence of the diverse physical processes involved.

\subsection{Variability indexes}
\label{sec:varindex}

\begin{figure}
	\centering
	\includegraphics[width=\columnwidth]{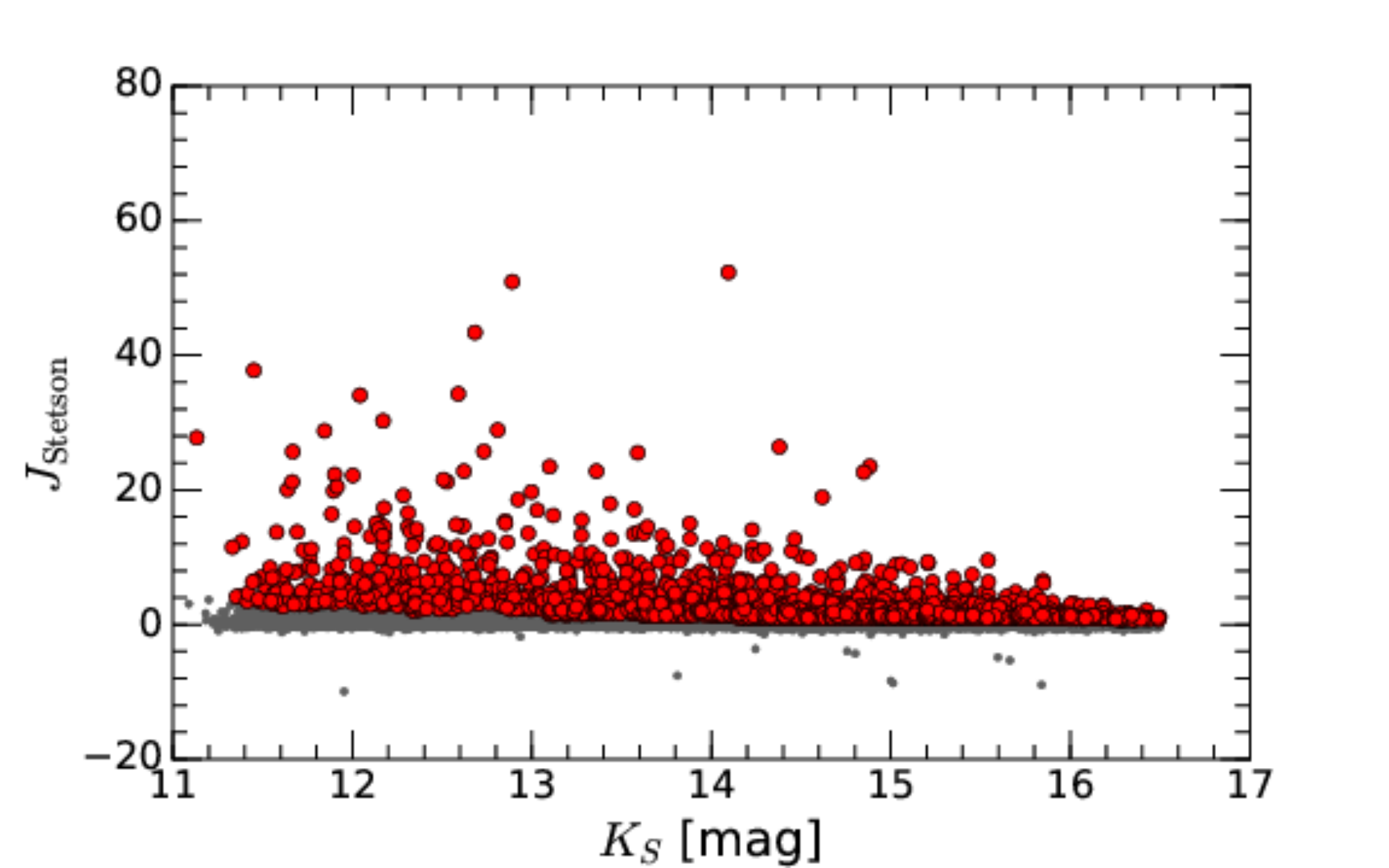}
	\caption{Stetson $J$ index versus mean magnitude for all 97,659 sources from this work. Red circles correspond to objects above $3\sigma$ for each magnitude bin. The sigma-clipping routine was repeated 50 times.
	\label{fig-2}}
\end{figure}

The variable selection was carried out differently than for \citetalias{nav16}. Given the plethora of variability indexes available in the literature \citep[see e.g.][]{soko17, ferr16}, we decided to choose the combination of indexes best suited for our data. The choice of variability indexes was similar to the selection carried out in \citet{bori16}: 

\begin{itemize}

    \item The mean variance $^{\sigma}/_{\mu}$, defined as the standard deviation over the mean magnitude of the light curve;

    \item The Stetson $J$ and $K$ indexes \citep{stet96}, adapted to single-band observations;

    \item The von Neumann ratio $\eta$ \citep{von41};

    \item The $\chi^2$ test.

\end{itemize}

The performance of each index was tested by counting the number of real variables selected and the number of false detections. Figure~\ref{fig-1} shows the test applied to all five indexes. It is clear that the variable population is more easily distinguished in some indexes. For each variability index, a sigma-clipping is applied for each bin of 0.5 magnitudes in order to select variable candidates. The value of the sigma-clipping is chosen so it minimizes the false detections and the number of missing real variables. This process is repeated until the number of candidates converges. This procedure was applied to the different variability indexes listed in the beginning of this section. Finally, we selected the combination of two indexes that gave the best ratio of false detections versus missed real variables. For this work, a combination of $^{\sigma}/_{\mu}$ and the Stetson $J$ index gives the most optimal results.


\section{Candidates selection}
\label{sec:candsel}

\begin{figure*}
\begin{center}
\begin{minipage}{0.5\textwidth}
  \centering
  \includegraphics[width=0.9\textwidth]{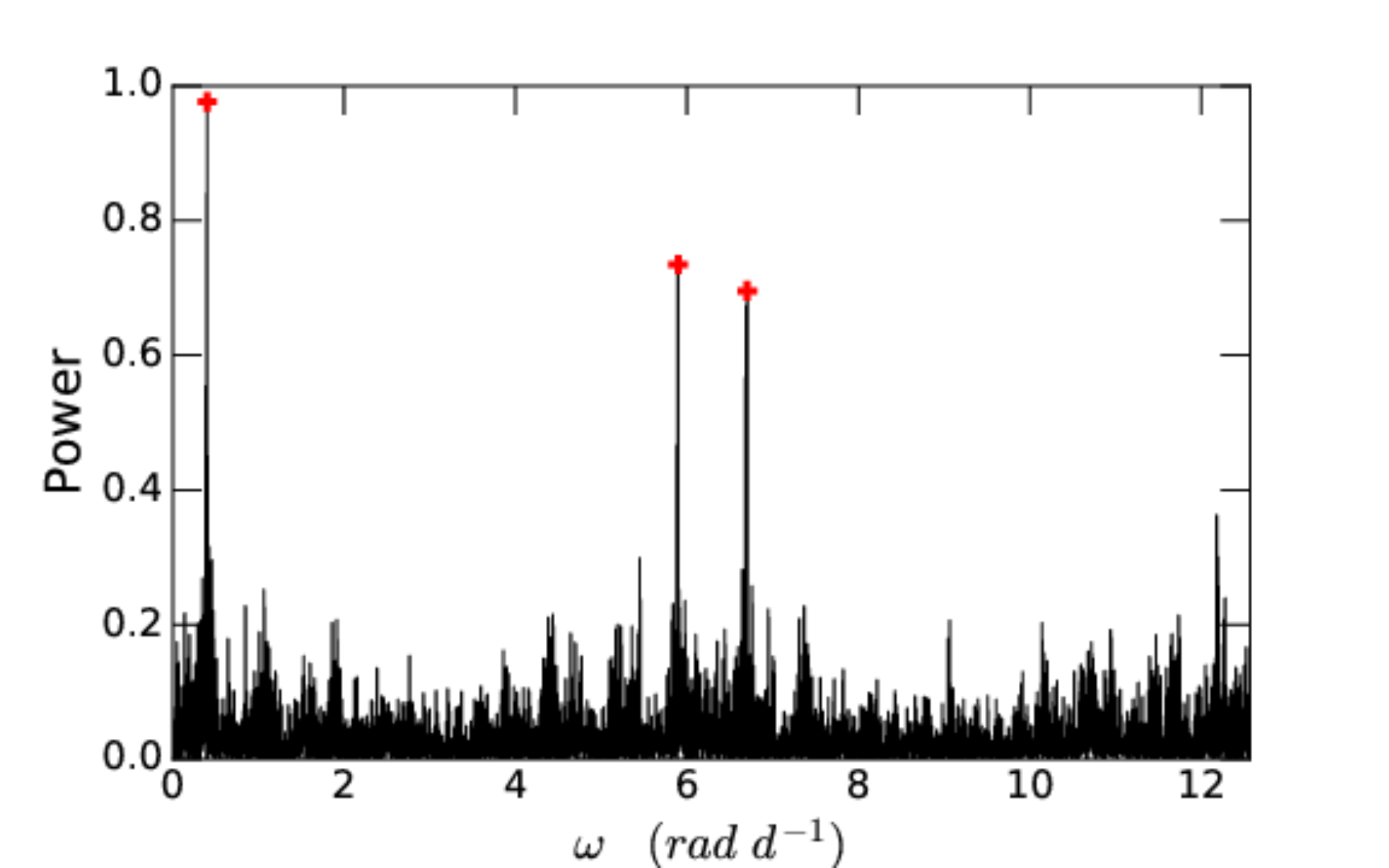}
\end{minipage}\hfill
\begin{minipage}{0.5\textwidth}
  \centering
  \includegraphics[width=0.9\textwidth]{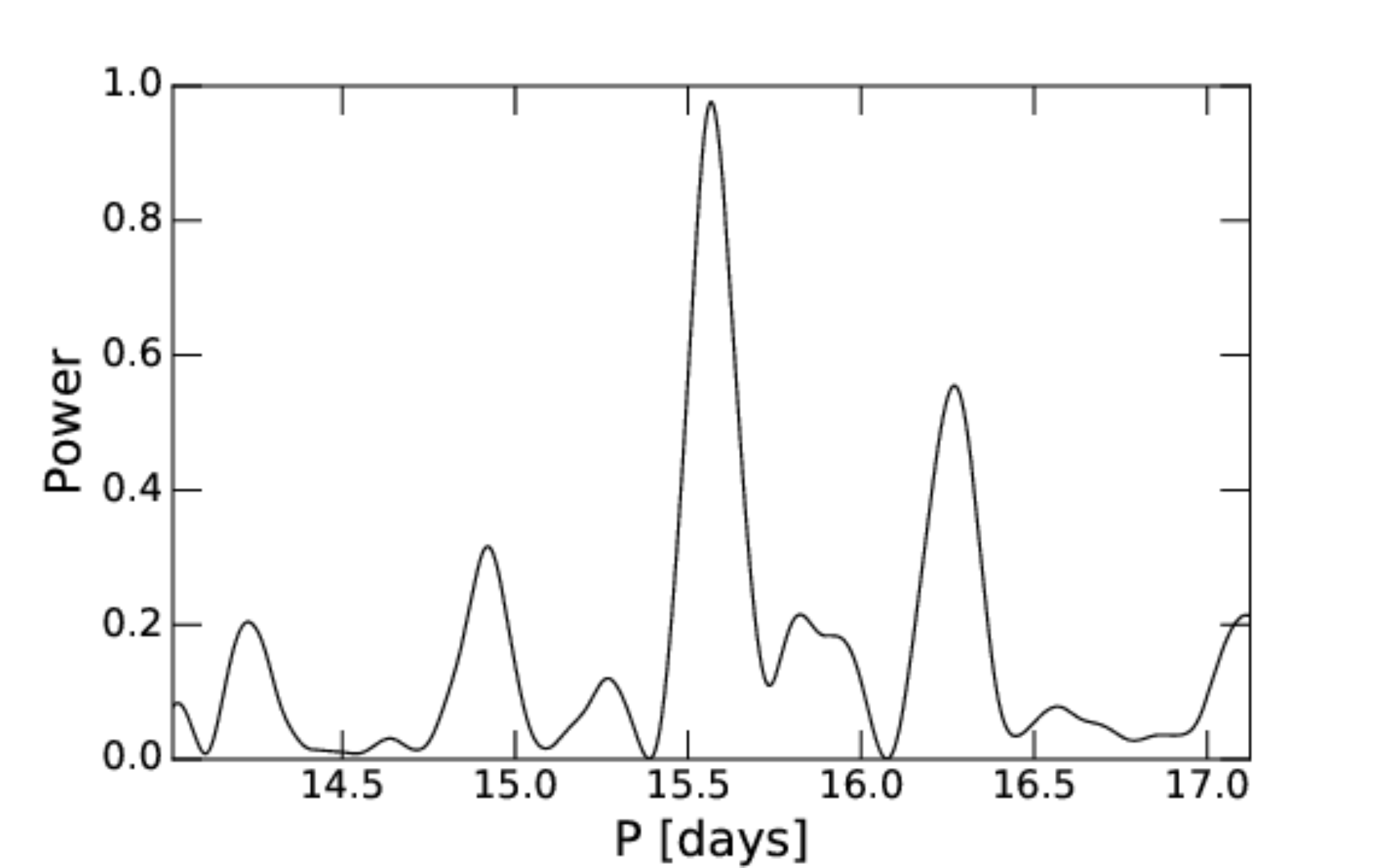}
\end{minipage}\hfill
\begin{minipage}{\textwidth}
  \centering
  \includegraphics[width=0.9\textwidth]{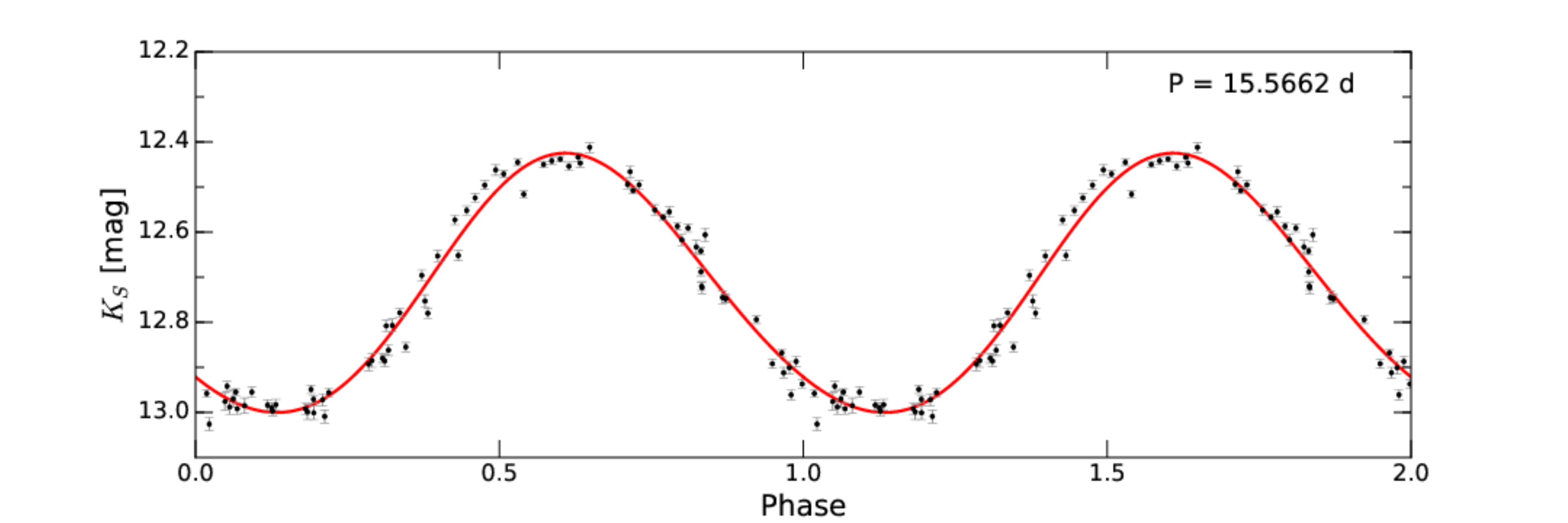}
\end{minipage}\hfill
\caption{(top left) GLS power spectrum for variable NV142. Red crosses mark the highest peaks in frequency space. (top right) Second iteration of the GLS periodogram, as explained in Section~\ref{sec:period}, now in period space. (bottom) Folded light curve for the period shown in the top right corner. The red line is a second-order Fourier fit to the light curve.\label{fig-3}}
\end{center}
\end{figure*}

To ensure that only good quality stars are used for the remaining analysis, we applied additional selection criteria:

\begin{enumerate}[label=(\roman*), align=parleft, leftmargin=*]

\item At least 25 epochs with good detections;

\item A mean magnitude of $K_S < 16.5$;

\item A peak-to-trough amplitude of $\Delta K_S > 0.2$.

\end{enumerate}

The minimum of 25 epochs is required to produce a well sampled light curve, in order to correctly discriminate the variability type, and to find a suitable period (if the variations show some regularity). We are aware that this cut will probably not detect transients like nova eruptions, but such phenomena are not the main scope of this paper. The mean-magnitude cut is used to exclude very faint objects for which photometric errors become comparable to the amplitudes of variation. The last cut takes into account the typical scatter found within VVV objects \citep{cont17a}.

After the cuts are applied, we calculate the two variability indexes discussed in the previous section. Objects with indexes above $3\sigma$ were chosen as potential candidates. Figure~\ref{fig-2} displays the results for the Stetson $J$ index. As we can infer from the graph, even at $3\sigma$ there is a large number of variable candidates. However, if we choose a larger sigma level, there is a risk that real variables could be missed. Nevertheless, the combination with the other variability index greatly reduces the number of candidates. In our case, 655 potential variable star candidates were found. A one-by-one visual inspection was performed in order to remove false detections produced by bad photometry, image artifacts, a nearby bright star or proximity with the image edges. The final number of variable star candidates after the above described conservative check is 353.

\subsection{Period estimation}
\label{sec:period}

Light curves were analyzed with the generalized Lomb-Scargle (GLS) periodogram \citep{lomb76,js82,zech09}, in order to find periodicity in our variables. We chose GLS over other period-finding algorithms, such as phase-dispersion minimization or analysis of variance, since it is more suited to work with irregularly sampled data, which is the case of our measurements \citep[e.g.,][and references therein]{vdp15}. For this, we selected an equally spaced frequency grid equivalent to a period range between 0.5 and 1800 days. The upper limit in the period is required, since we cannot study temporal variations with timescales longer than the total span of the VVV observations, which corresponds to 1861 days. The most prominent peaks of the periodogram were analyzed to remove possible aliases. To distinguish real periods from irrelevant ones, we arbitrarily set a lower peak value of 0.6 in the corresponding power spectrum, plus a false alarm probability (FAP) inferior than 1\%. After this step, we refined the period search, this time adjusting the frequency grid to be between values equivalent to periods around 10\% from the first guess. The entire process is summarized in Figure~\ref{fig-3}.


\section{Variable stars catalog}
\label{sec:catalog}

\begin{figure*}
	\centering
	\includegraphics[width=0.9\textwidth]{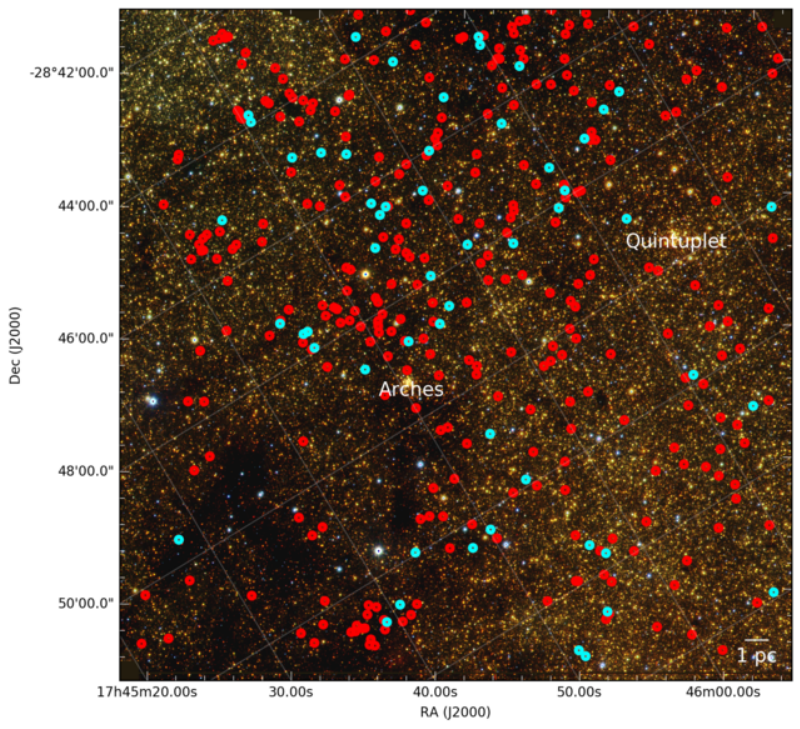}
	\caption{$JHK_S$ composite image of the $11.5'\times11.5'$ region belonging to the VVV tile b333. This area is equivalent to a single chip of the VIRCAM camera. Location of variable stars is marked as red circles, and those with literature counterparts are shown with cyan circles. The positions of Arches and Quintuplet are marked in the image. The segment in the lower-right corner marks a length of 1~pc at the distance of the GC.\label{fig-4}}
\end{figure*}

\subsection{General properties}
\label{sec:properties}

Figure~\ref{fig-4} displays the $JHK_S$ composite image of the field studied in this work. The 299 variable stars with no counterparts are represented as red circles; cyan circles correspond to the 54 stars with literature counterparts. The majority of the variables are located in the upper half of the image, and a noticeable void  is present in the bottom left quadrant. This is coincident with a dark cloud that is observed in the image. The 33 variable stars near Quintuplet studied in \citetalias{nav16} are not shown. 

For each variable star found, we queried the Vizier database \citep*{ochs00} and searched for objects within a 1\arcsec\ radius, in order to check if our star had already been detected by previous variability surveys (see more details in Section~\ref{sec:literature}), as well as to find additional photometric information (for example, WISE measurements).

Furthermore, we cross-matched our variable stars with the point-source catalog of \citet{rami08}. The authors used the {\it Spitzer Space Telescope}, equipped with the Infrared Array Camera (IRAC), to survey the inner $2.0\degree \times 1.4\degree$ region of the GC. The array has four channels, 1 through 4, equivalent to the [3.6], [4.5], [5.8] and [8.0]~$\micron$ bands, respectively. All of our variable stars were matched with their closest counterpart from the \citeauthor{rami08} sources, within 1\arcsec. There are 182 stars with at least one mid-IR measurement in the \citeauthor{rami08} catalog.

The full summary with the main properties of the 353 candidate variable stars is presented in Table~\ref{tbl:tbl-1}. Column 1 shows the name assigned to each variable. Columns 2 and 3 display the coordinates of each object, while columns 4 and 5 show the corresponding Galactic coordinates. Columns 6-8 present the nearly simultaneous $J$, $H$ and $K_S$ VVV magnitudes from 2010. Columns 9-12 contain magnitudes from the four {\it Spitzer} bands, taken from \citet{rami08}. Column 13 corresponds to the mean magnitude, defined simply as the mean value between the maximum and minimum magnitudes for each star. Column 14 displays the peak-to-trough amplitude of the variation. Column 15 shows the period obtained using the procedure described in Section~\ref{sec:period}. Column 16 presents the variability type and, finally, column 17 includes the reference for the variables found in previous works.

\begin{landscape}
	\begin{table}
	\caption{Summary of the main properties of the first 40 variable stars. See Section~\ref{sec:properties} for a description of each column. The full version of this table is available online (see Supporting Information).}
    \label{tbl:tbl-1}
    \begin{tabular}{lcccccccccccccccc}
    \hline
    Name & RA & Dec. & $l$ & $b$ & $J$ & $H$ & $K_S$ & $[3.6]$ & $[4.5]$ & $[5.8]$ & $[8.0]$ & $\langle K_S \rangle$ & $\Delta K_S$ & Period & Class & Lit. ref.$^a$ \\
 & (J2000) & (J2000) & (\degree) & (\degree) & (mag) & (mag) & (mag) & (mag) & (mag) & (mag) & (mag) & (mag) & (mag) & (days) & & \\
	\hline
NV001 & 17:45:21.16 & -28:50:48.36 & 0.045810 & 0.096523 & -- & 18.13 & 14.25 & -- & -- & -- & -- & 14.16 & 0.97 & 1248 & SR & -- \\
NV002 & 17:45:23.25 & -28:50:58.59 & 0.047359 & 0.088541 & 19.03 & 14.55 & 12.03 & 10.52 & 10.1 & 9.68 & 8.99 & 12.23 & 0.4 & -- & IRR & -- \\
NV003 & 17:45:23.46 & -28:50:06.46 & 0.060115 & 0.095437 & -- & 18.44 & 14.76 & -- & 11.62 & 11.45 & -- & 14.54 & 1.11 & 1144 & SR & -- \\
NV004 & 17:45:27.15 & -28:50:17.78 & 0.064462 & 0.082291 & -- & -- & 16.39 & -- & -- & -- & -- & 16.19 & 1.29 & -- & Eruptive & -- \\
NV005 & 17:45:28.11 & -28:49:34.76 & 0.076481 & 0.085539 & -- & 17.15 & 12.28 & 8.9 & 8.14 & -- & -- & 12.0 & 1.06 & 498 & Mira & 1 \\
NV006 & 17:45:30.82 & -28:51:06.02 & 0.060007 & 0.063877 & -- & 17.45 & 13.37 & -- & -- & -- & -- & 13.73 & 1.24 & -- & LPV & -- \\
NV007 & 17:45:32.07 & -28:48:40.54 & 0.096883 & 0.081029 & -- & 14.68 & 11.97 & 10.35 & 10.01 & -- & 9.67 & 12.02 & 0.46 & -- & IRR & -- \\
NV008 & 17:45:32.66 & -28:52:07.61 & 0.048898 & 0.049242 & -- & 16.21 & 13.23 & -- & 10.97 & 10.29 & -- & 13.44 & 0.84 & 1502 & SR & -- \\
NV009 & 17:45:33.19 & -28:52:23.69 & 0.046093 & 0.045266 & -- & 15.99 & 13.07 & 10.96 & 10.45 & 9.49 & -- & 13.34 & 0.51 & -- & LPV & -- \\
NV010 & 17:45:33.75 & -28:48:36.44 & 0.101040 & 0.076409 & -- & 15.75 & 13.48 & 12.52 & 12.52 & -- & -- & 13.6 & 0.5 & -- & LPV & -- \\
NV011 & 17:45:34.53 & -28:47:34.48 & 0.117224 & 0.082929 & -- & 16.89 & 12.29 & 10.97 & 9.51 & 9.05 & -- & 12.8 & 1.03 & -- & IRR & -- \\
NV012 & 17:45:34.57 & -28:52:11.92 & 0.051510 & 0.042669 & -- & 15.68 & 12.48 & 10.57 & 10.13 & 9.2 & -- & 12.46 & 0.42 & 1738 & SR & -- \\
NV013 & 17:45:35.65 & -28:47:43.67 & 0.117173 & 0.078115 & -- & 16.05 & 13.54 & 12.18 & 11.6 & -- & -- & 13.62 & 0.47 & -- & LPV & -- \\
NV014 & 17:45:35.67 & -28:51:51.48 & 0.058450 & 0.042200 & -- & 15.06 & 12.24 & -- & -- & -- & -- & 12.31 & 0.48 & 1430 & SR & -- \\
NV015 & 17:45:36.19 & -28:52:34.46 & 0.049252 & 0.034350 & -- & 18.0 & 14.86 & -- & -- & -- & -- & 15.35 & 1.22 & -- & IRR & -- \\
NV016 & 17:45:36.52 & -28:52:36.19 & 0.049477 & 0.033063 & -- & 15.82 & 12.21 & 10.66 & 9.83 & 9.47 & -- & 12.59 & 0.75 & -- & IRR & -- \\
NV017 & 17:45:36.85 & -28:52:31.56 & 0.051192 & 0.032721 & -- & 15.37 & 11.90 & 10.06 & 9.46 & 8.21 & -- & 11.88 & 0.46 & 1003 & SR & -- \\
NV018 & 17:45:36.93 & -28:52:56.19 & 0.045514 & 0.028888 & -- & -- & 13.91 & 12.82 & 11.1 & -- & -- & 14.88 & 3.47 & -- & RCB? & -- \\
NV019 & 17:45:37.16 & -28:52:37.58 & 0.050355 & 0.030883 & -- & 18.06 & 14.65 & -- & -- & -- & -- & 14.86 & 0.86 & -- & IRR & -- \\
NV020 & 17:45:37.25 & -28:52:51.92 & 0.047123 & 0.028530 & -- & 15.72 & 11.74 & 10.47 & 9.21 & 8.25 & 7.44 & 11.9 & 0.81 & -- & IRR & -- \\
NV021 & 17:45:37.27 & -28:52:59.88 & 0.045281 & 0.027304 & -- & -- & 16.68 & -- & -- & -- & -- & 15.19 & 1.47 & -- & Eruptive & -- \\
NV022 & 17:45:37.30 & -28:52:38.45 & 0.050428 & 0.030301 & -- & -- & 16.07 & -- & -- & -- & -- & 15.12 & 0.98 & -- & Dipper & -- \\
NV023 & 17:45:37.36 & -28:50:21.42 & 0.083019 & 0.049961 & -- & 16.43 & 13.15 & -- & -- & -- & -- & 13.01 & 0.38 & 259 & SR & -- \\
NV024 & 17:45:37.50 & -28:46:55.46 & 0.132114 & 0.079342 & -- & 15.16 & 12.57 & 11.42 & 11.04 & 10.01 & -- & 12.73 & 0.61 & -- & LPV & -- \\
NV025 & 17:45:37.54 & -28:50:44.99 & 0.077767 & 0.046002 & -- & 17.20 & 13.52 & -- & 11.06 & 10.61 & -- & 13.32 & 0.44 & -- & IRR & -- \\
NV026 & 17:45:38.05 & -28:52:27.59 & 0.054412 & 0.029561 & -- & 17.68 & 14.22 & -- & -- & -- & -- & 14.02 & 0.87 & -- & LPV & -- \\
NV027 & 17:45:38.53 & -28:52:19.61 & 0.057222 & 0.029214 & -- & 18.07 & 15.22 & -- & -- & -- & -- & 15.51 & 0.94 & -- & Dipper & -- \\
NV028 & 17:45:38.60 & -28:50:43.11 & 0.080243 & 0.042948 & -- & 16.51 & 12.36 & 11.16 & 9.83 & 9.43 & -- & 12.21 & 0.48 & -- & LPV & -- \\
NV029 & 17:45:38.65 & -28:52:50.89 & 0.050035 & 0.024310 & -- & -- & 14.18 & -- & 11.35 & -- & -- & 14.61 & 1.03 & -- & LPV & -- \\
NV030 & 17:45:38.97 & -28:52:41.55 & 0.052849 & 0.024677 & -- & 15.71 & 12.37 & 10.85 & -- & -- & -- & 12.92 & 1.0 & 1382 & SR & -- \\
NV031 & 17:45:38.99 & -28:52:25.70 & 0.056656 & 0.026893 & -- & 17.45 & 13.71 & 13.08 & 11.54 & -- & -- & 13.78 & 1.17 & -- & IRR & -- \\
NV032 & 17:45:39.09 & -28:52:45.14 & 0.052229 & 0.023779 & -- & 17.04 & 12.30 & 9.96 & 8.32 & 7.72 & 7.52 & 12.15 & 1.09 & -- & IRR & 1 \\
NV033 & 17:45:40.16 & -28:46:51.91 & 0.138030 & 0.071543 & 18.18 & 14.81 & 12.48 & 10.14 & 9.35 & 8.76 & -- & 12.51 & 0.6 & 311 & SR & -- \\
NV034 & 17:45:40.21 & -28:52:53.49 & 0.052388 & 0.019067 & -- & 16.51 & 12.95 & -- & -- & -- & -- & 12.84 & 0.24 & -- & IRR & -- \\
NV035 & 17:45:40.70 & -28:45:27.64 & 0.159028 & 0.082069 & -- & -- & 16.09 & -- & 12.65 & -- & -- & 15.54 & 1.79 & -- & Eruptive/FUor & -- \\
NV036 & 17:45:40.74 & -28:52:36.65 & 0.057376 & 0.019874 & 17.71 & 14.82 & 13.13 & -- & -- & -- & -- & 13.16 & 0.45 & 0.5564 & EB & 2 \\
NV037 & 17:45:40.85 & -28:49:14.67 & 0.105486 & 0.048742 & -- & 18.08 & 15.77 & -- & -- & -- & -- & 15.65 & 1.48 & -- & Eruptive/EXor & -- \\
NV038 & 17:45:41.06 & -28:44:22.79 & 0.175093 & 0.090326 & -- & 17.25 & 15.31 & -- & -- & -- & -- & 15.54 & 1.06 & -- & Fader & -- \\
NV039 & 17:45:41.07 & -28:52:51.98 & 0.054384 & 0.016604 & -- & 16.32 & 12.45 & 9.93 & 9.18 & 8.75 & -- & 12.66 & 0.48 & -- & IRR & -- \\
NV040 & 17:45:41.62 & -28:45:04.70 & 0.166226 & 0.082507 & -- & -- & -- & -- & -- & -- & -- & 16.11 & 1.01 & -- & Eruptive & -- \\

	\hline
	\multicolumn{13}{l}{$^a$ {\bf References.} (1) \citetalias{mkn09}; (2) \citetalias{mfk13}; (3) \citet{woo98}.}
    \end{tabular}
\end{table}
\end{landscape}

\begin{figure*}
\centering
\includegraphics[width=\textwidth]{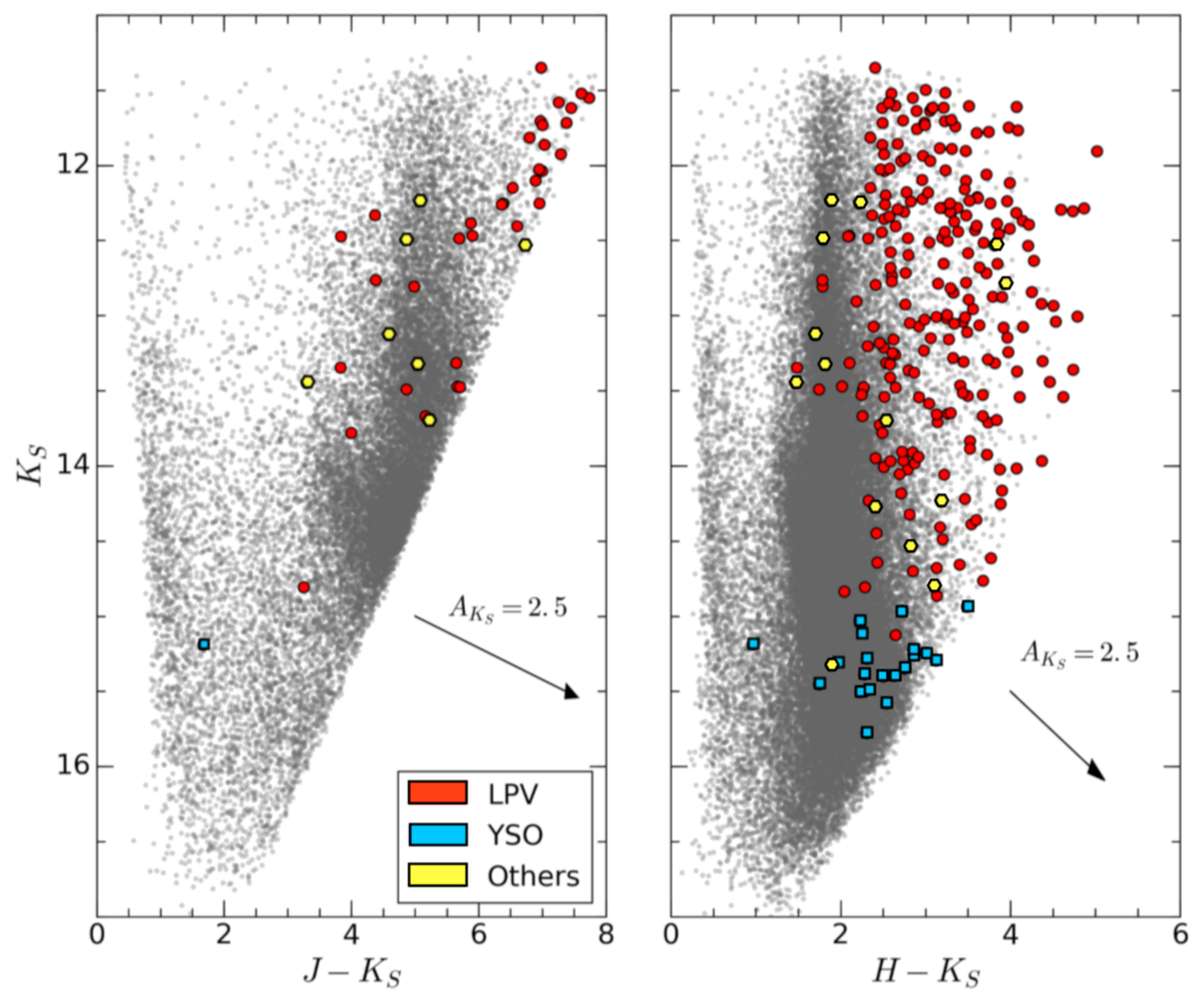}
\caption{CMDs for all the sources detected in our region: (left) $K_S$ versus $(J-K_S)$ and (right) $K_S$ versus $(H-K_S)$. In both plots, non-variable objects are represented by grey dots. Coloured symbols correspond to different variability types: red circles, blue squares and yellow hexagons are red giant stars (both pulsating LPVs and other giants with long-term trends), YSO candidates and other variable classes, respectively. Arrows represent reddening vectors for an extinction of $A_{K_S}=2.5$ mag. All magnitudes, including those of variable stars, correspond to the first epoch of observations in 2010.
\label{fig-5}}
\end{figure*}


\subsection{Colour-magnitude diagram}
\label{sec:cmd}

Colour-magnitude diagrams (CMDs) for the entire $11.5'\times11.5'$ region are shown in Figure~\ref{fig-5}. Non variable stars are displayed as grey dots. In addition, reddening vectors representing an extinction of $A_{K_S}=2.5$~mag, according with the  \citet[][hereafter AMC17]{alon17} law, are shown as reference. For an in-depth discussion of the different extinction laws towards the GC, please see Section~\ref{sec:dist}.

For clarity purposes, we separated variable classes into three large groups: pulsating red stars and red stars with no clear classification, YSO candidates, and other variable classes (such as type II Cepheids and EBs). This corresponds to red, blue, and yellow symbols in the figure, respectively.

In the left panel of Figure~\ref{fig-5} we have a $K_S$ versus $(J-K_S)$ CMD. Since most of our catalog members are intrinsically red sources, they are often too faint in the $J$-band to be detected by our photometry. About 15\% of our sources have $J$-band photometric data; of these, we only have 41 variables with simultaneous $J$, $H$ and $K_S$ magnitudes in the 2010 epoch. Similar numbers are found for the 2015 epoch. The position of the red circles is consistent with these stars being luminous AGB stars located near the GC. Those variable sources with bluer colours are probably foreground objects in the red giant branch (RGB), affected by a less severe interstellar extinction.

The right panel shows a $K_S$ versus $(H-K_S)$ CMD. In this case, a larger number of variable stars is present. There are 266 sources, i.e., $\sim75\%$, that have both $H$ and $K_S$ magnitudes in the first epoch. Red giant stars present a wide range of colours, which is primarily caused by the different levels of extinction (both interstellar and circumstellar) that affect them.


\subsection{Colour-colour diagram}
\label{sec:ccd}

\begin{figure}
\centering
\includegraphics[width=\columnwidth]{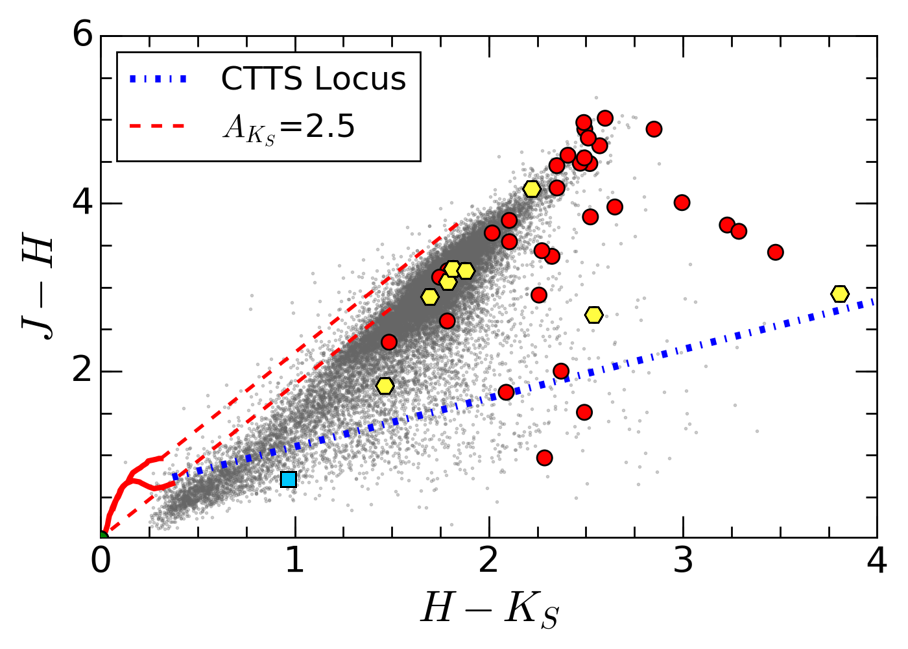}
\caption{$(H-K_S)-(J-H)$ colour-colour diagram. Grey dots are non-variable sources. Large symbols correspond to variable stars, and their colours are the same as in Figure~\ref{fig-5}. Reddening vectors corresponding to $A_{K_S}=2.5$ are plotted as dashed lines. Solid lines mark the unreddened location of dwarfs and giants, extracted from \citet{bess88}. The blue dot-dashed line is the CTTS locus described in Meyer et al. (1997). \label{fig-6}}
\end{figure}


The colour-colour diagram for the 2010 epoch is presented in Figure~\ref{fig-6}. As discussed in Section~\ref{sec:cmd}, there are 41 variable stars with magnitudes in these three bands. Solid lines correspond to \citet{bess88} magnitudes of dwarfs and giants. For reference, we draw reddening vectors equivalent to an extinction of $A_{K_S} = 2.5$ mag (AMC17 extinction law).

There is a group of LPVs located at the tip of the elongated AGB, as expected, while others are shown as highly reddened stars, an indication that they are surrounded by thick dust envelopes. The rightmost point of Figure~\ref{fig-6} corresponds to NV235, an object that undergoes a systematic decline of $\sim2$ mag, and that is classified as a candidate RCB star. On the other hand, there are three sources below the CTTS line, namely NV146, NV195 and NV334. The first one is labeled as an YSO candidate, and will be discussed later. The other two are classified as red giant stars, though their colours are not consistent with the RGB and AGB of stars in the GC. The mean magnitudes of these sources, 14.45 and 14.02~mag, respectively, could indicate that they are faint red stars located in zones with a lower interstellar extinction. It is also possible that these stars are foreground red dwarfs. Unfortunately, the stars lack Gaia DR2 (Gaia Collaboration, 2018) measurements within 1 arcsec radius, thus we can not check their proper motion.  



\subsection{Mid-infrared properties}
\label{sec:mid-ir}

\begin{figure*}
\begin{center}
\begin{minipage}{0.5\textwidth}
  \centering
  \includegraphics[width=\textwidth]{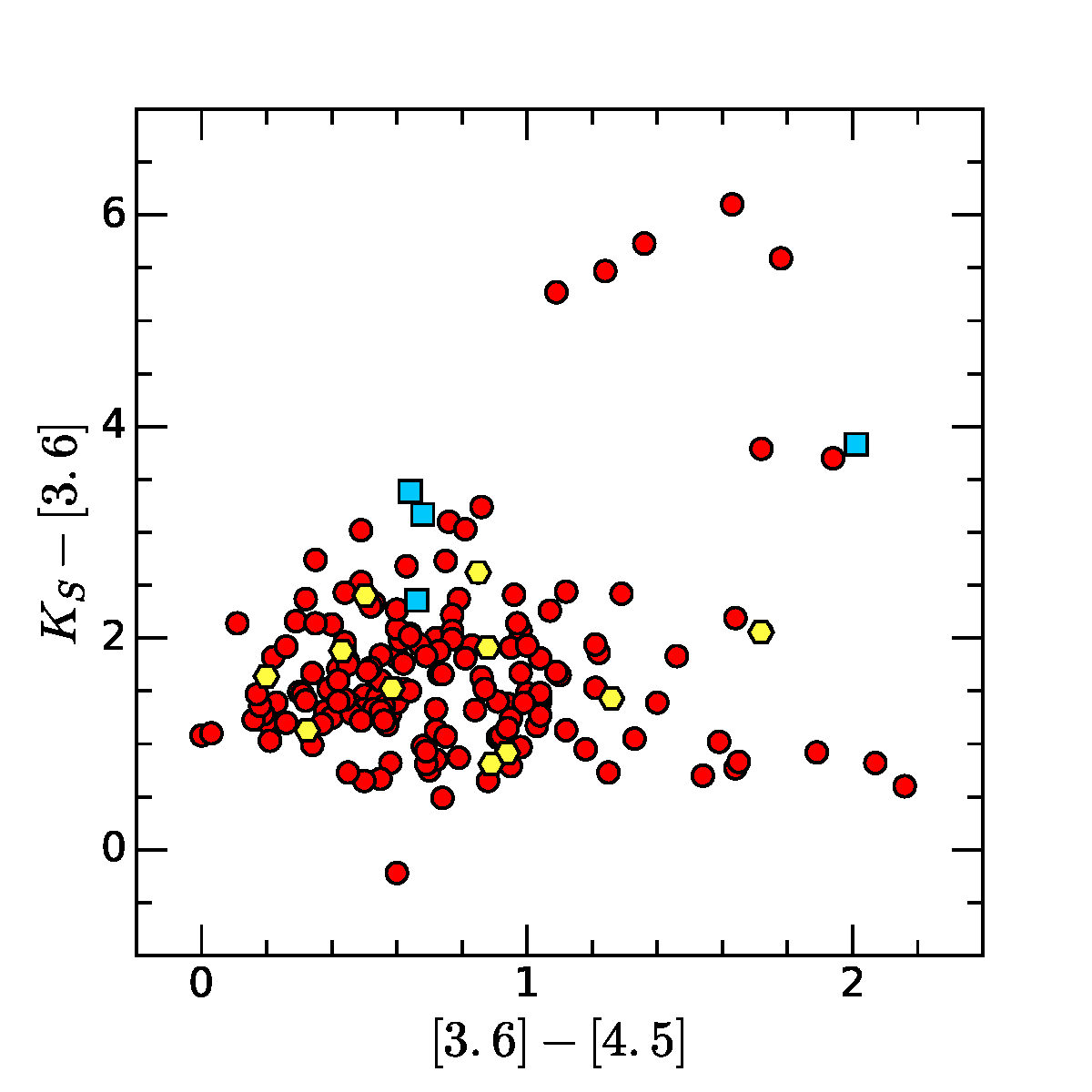}
\end{minipage}\hfill
\begin{minipage}{0.5\textwidth}
  \centering
  \includegraphics[width=\textwidth]{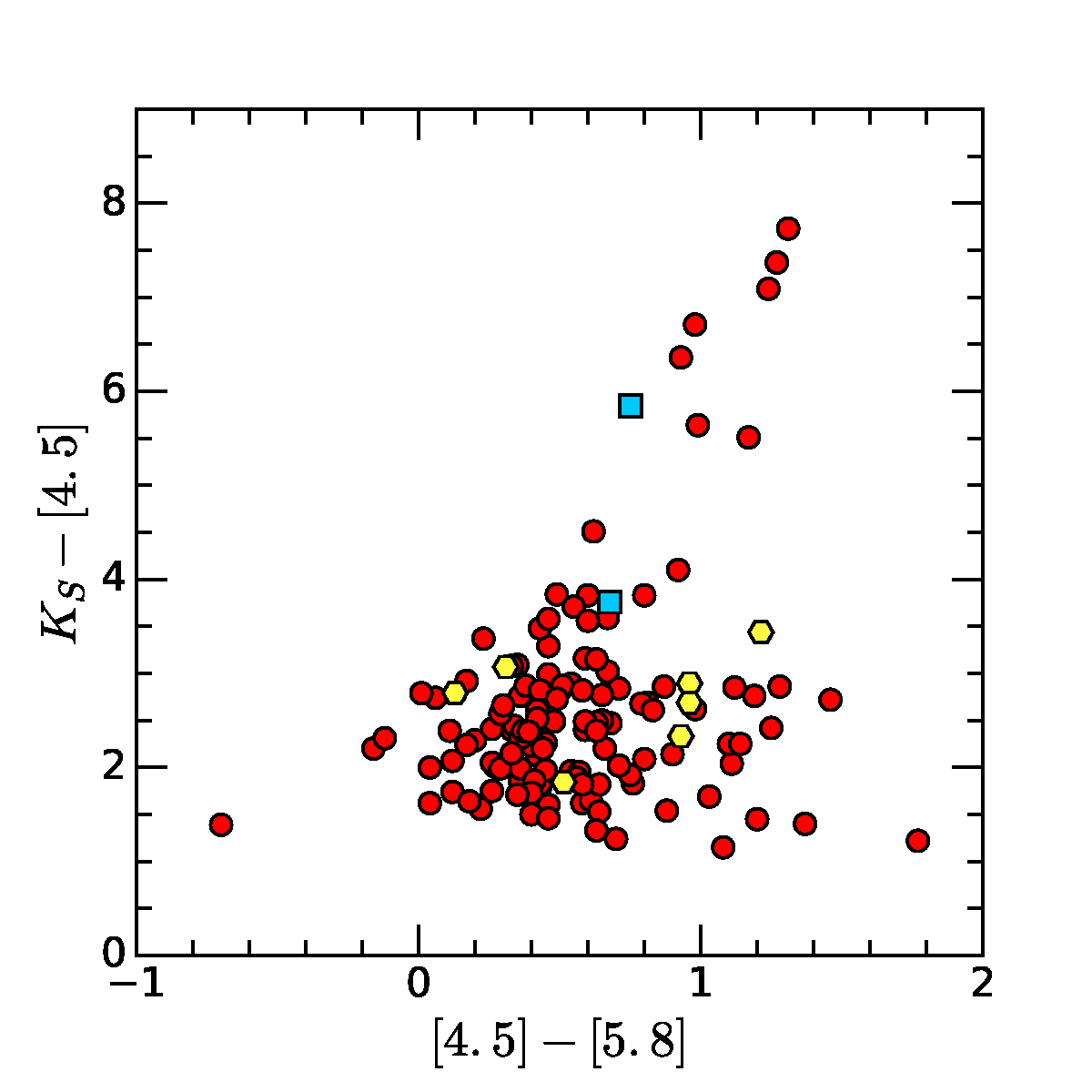}
\end{minipage}\hfill
\caption{(left) $(K_S-[3.6]),([3.6]-[4.5])$ colour-colour diagram for variable stars with {\it Spitzer} magnitudes. (right) $(K_S-[4.5]),([4.5]-[5.8])$ colour-colour diagram. The colours and symbols of variable stars are the same as in Figure~\ref{fig-5}.\label{fig-8}}
\end{center}
\end{figure*}

The {\it Spitzer} magnitudes available in the \citet{rami08} catalog allowed us to analyze the mid-IR behavior of our variable stars. The total numbers of counterparts found in the {\it Spitzer} $[3.6]$, $[4.5]$, $[5.8]$ and $[8.0]$ bands were 182 ($51.6\%$), 206 ($58.4\%$), 149 ($42.2\%$) and 34 ($9.6\%$), respectively. 

Figure~\ref{fig-8} presents the $(K_S-[3.6])$, $([3.6]-[4.5])$ and the $(K_S-[4.5])$, $([4.5]-[5.8])$ colour-colour diagrams. In both panels, the majority of our variable stars are distributed within a relatively narrow range of colours. There is a small group of sources that form a ``tail'' in the upper-right section of both plots, although it is more noticeable in the right panel. All of them but one correspond to stars near the tip of the AGB that have large mass-loss rates, as inferred from the amount of infrared excesses they present. For example, the star with the largest $(K_S-[4.5])$ colour is NV118, an infrared Mira with an extremely long period of 937~d, and a $K_S$ amplitude of 3.18~mag. According to \citet{wood07}, this implies a mass-loss rate of $\sim10^{-5}\msun\,$yr$^{-1}$, consistent with the expectations for luminous AGB stars \citep[e.g.,][]{gs18,ho18}.

The exception in this group of stars with red $(K_S-[4.5])$ colours is NV303, which is classified as a YSO fader candidate, since its light curve shows a systematic decline ($\sim1.5$ mag).  However, the large and relatively fast brightness decline also resembles an RCB variable. The faintness of this star ($\langle K_S \rangle = 15.36$) makes it hard to obtain a proper classification.

\begin{figure*}
\includegraphics[width=8cm]{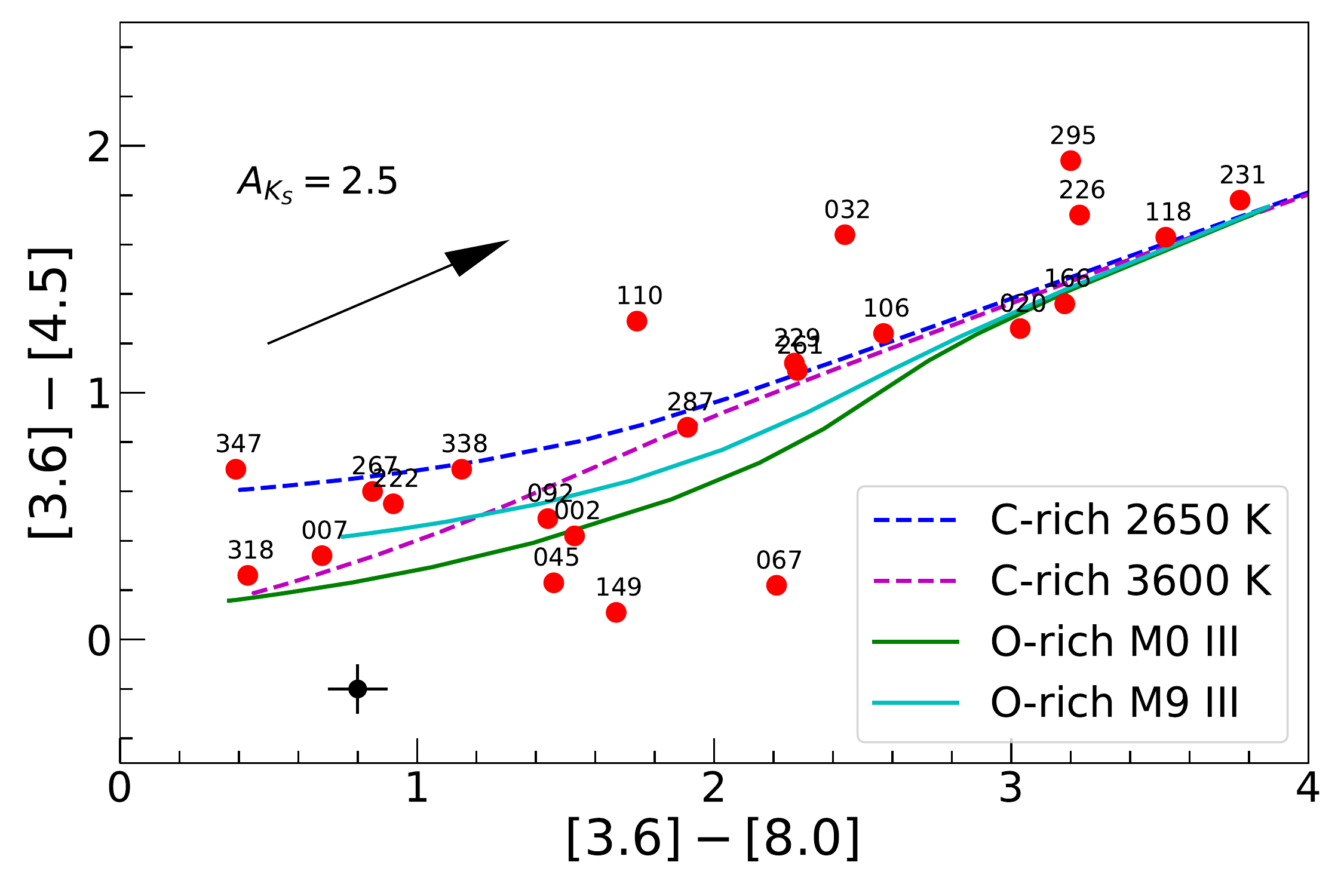}
\includegraphics[width=8cm]{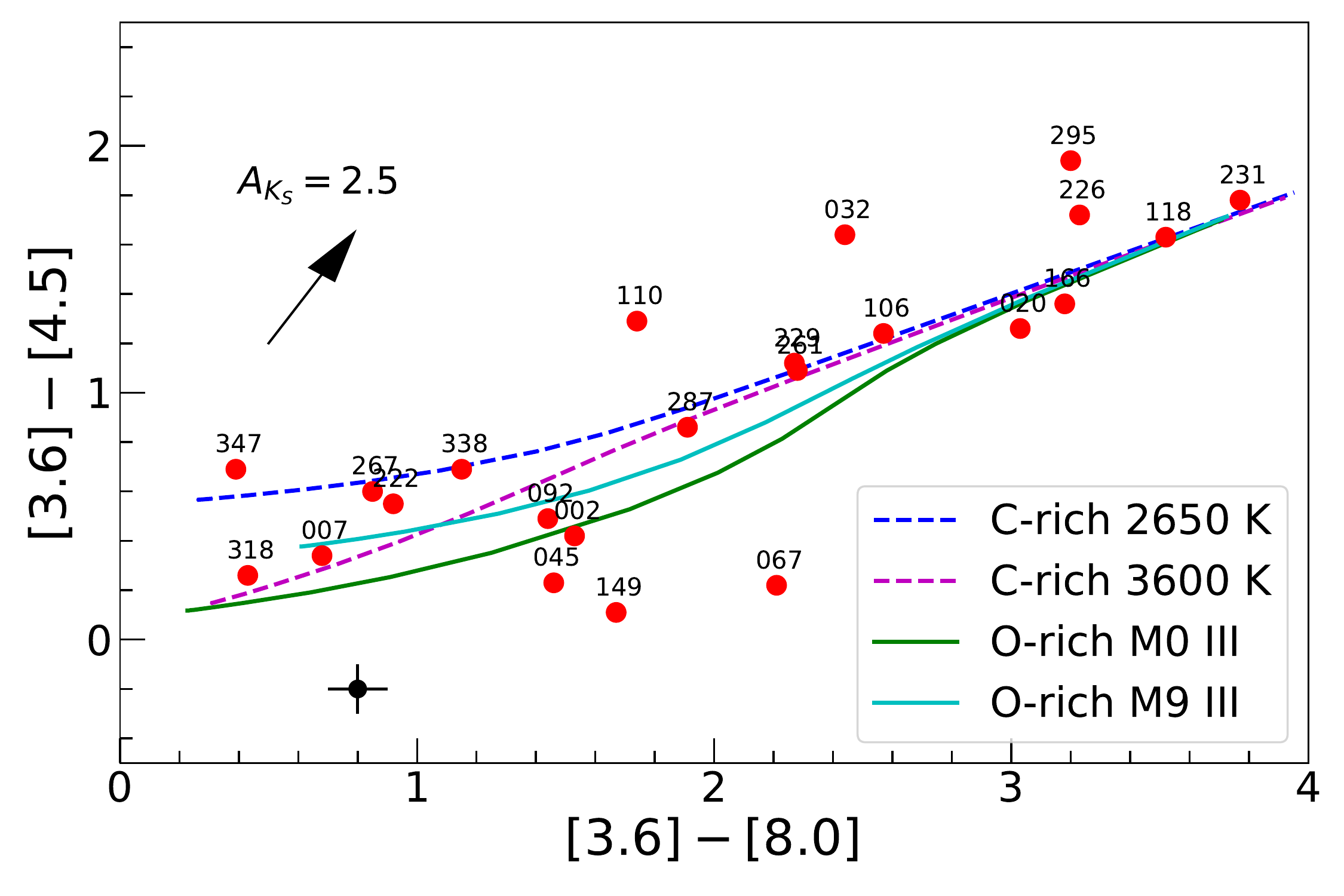}
\caption{$([3.6]-[4.5])$, $([3.6]-[8.0])$ colour-colour diagram. All tracks are synthetic models of mid-IR colours of AGB stars from \citet{groe06}, shifted assuming an average extinction of $A_{K_S}=2.5$ and the AMC17 extinction law (upper part), and the \citetalias{nish09}. Solid lines correspond to O-rich stars with silicate dust of spectral classes M0\,III (green) and M9\,III (cyan). Dashed lines are C-rich stars with amorphous carbon envelopes and effective temperatures of $T_{\rm eff}=2650$~K (blue) and $T_{\rm eff}=3600$~K (magenta). The number ID for each variable star in the diagram is also shown for easier identification. The arrow represents a reddening vector for $A_{K_S}=2.5$~mag. The black circle shows the average error bar sizes for these colours.\label{fig-10}}
\end{figure*}

A comparison with dust radiative transfer models for AGB stars from \citet{groe06} in a $([3.6]-[4.5]),([3.6]-[8.0])$ colour-colour diagram is shown in Figure~\ref{fig-10}. The 23 sources shown correspond to AGB variables. Models for C-rich and O-rich stars for different effective temperatures are included. The mid-IR colours from \citet{rami08} are not corrected for reddening. If we assume a mean value of $A_{K_S}=2.5$~mag for the extinction in the region, we have, according to \citetalias{nish09}, $E_{[3.6]-[4.5]}=0.28$~mag and $E_{[3.6]-[8.0]}=0.18$~mag. Models were shifted using these reddening values. Since extinction in these IRAC bands is nearly half the value in the $K_S$-band, even an increase of 0.5 mag in $A_{K_S}$ would lead to a shift of $\sim0.05$ mag in colour for the model tracks. The same calculations were performed using the \citet{alon17} extinction law and, as can be seen from Figure~\ref{fig-10}, this is within the order of the photometric errors of the stars' mid-IR magnitudes. While \citet{groe06} provides additional models for alternative dust compositions, the differences with the tracks illustrated are smaller than the colour errors, so they are not included in this discussion.

Figure~\ref{fig-10} shows that, for $[3.6]-[8.0]\lesssim3$ mag, the C-rich and O-rich tracks are separated enough to distinguish both surface chemistry. There are at least six variables, NV229, NV261, NV267, NV287, NV338 and NV347, that have mid-IR colours consistent with carbon stars. All of them are classified as IRR, except for NV229 and NV338 that are SR variables. On the other hand, NV002 and NV092 have colours corresponding to M0\,III stars. Both are IRR variables.

There are four Mira variables with confirmed OH emission in our catalog, namely, NV118, NV166, NV231 and NV295. All of them have large $[3.6]-[8.0]$ excesses, and, with the exception of NV295, they lie very close to the theoretical tracks. For $[3.6]-[8.0]<1$ mag, the O-rich tracks become closer to the 3600 K C-rich track, so it is not clear to which group NV007 and NV318 belong.  

The above analysis, although it does not replace the need of spectroscopy to derive the photospheric composition of these AGB stars, still provides a reasonable first-order estimate. \citet{mare08} used the same models from \citet{groe06} to compare them with the colour distribution of AGB stars observed with {\it Spitzer} and with known C/O ratios. The authors find that carbon stars clearly separate from the O-rich ones, since the former have systematically redder $[3.6]-[4.5]$ colours than the later, for $[3.6]-[8.0]>1$ mag. This is caused by carbon stars lacking important dust features in the IRAC bands. Hence, C-rich stars have smaller $[3.6]-[8.0]$ excess for a given $[3.6]-[4.5]$ colour, which makes C-rich stars appear above O-rich stars in the colour-colour diagram \citep{mare08}, as indeed seen in Figure~\ref{fig-10}.

\section{Literature counterparts}
\label{sec:literature}

\begin{figure}
\centering
\includegraphics[width=\columnwidth]{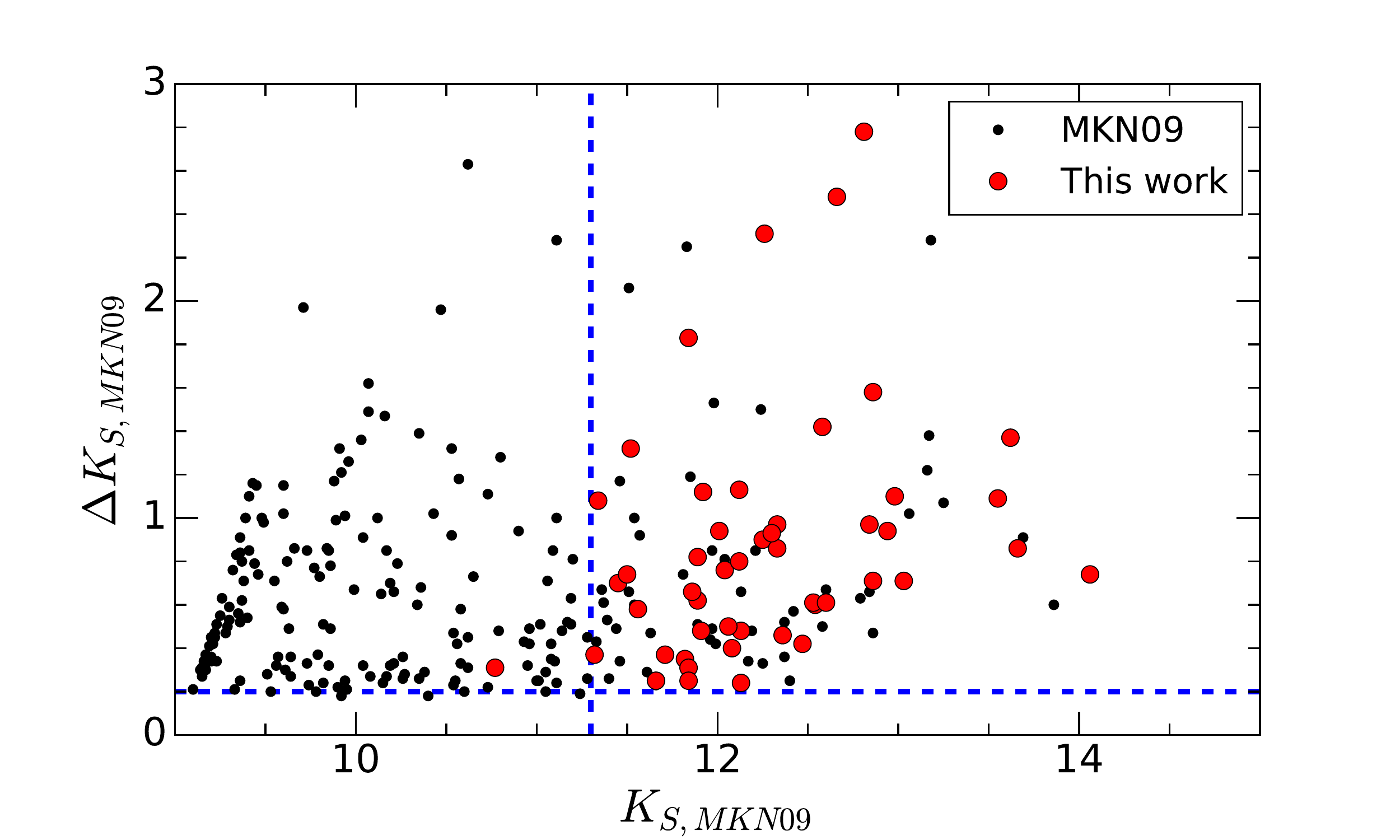}
\caption{Distribution of \citetalias{mkn09} $K_S$ amplitude versus mean magnitude for all sources with $mflag=0$ in our field (black circles). Blue dashed vertical and horizontal lines represent $\bar{K}_S=11.5$~mag and $\Delta K_S=0.2$~mag, respectively (see text for details). Variable stars of our present work with \citetalias{mkn09} counterparts are represented as red circles.\label{fig-11}}
\end{figure}

As mentioned before, there is an overlap between our present work and 
previous variability studies in the region. In order to find counterparts, we 
performed cross-matches with the variables of \citetalias{woo98} and \citet{
mkn09,mfk13}. For all these cases, the tolerance was set at 1\arcsec. There 
are 9 \citetalias{woo98} stars in our field. Of these we have only two 
counterparts, NV118 and NV231. Regarding the other seven, five of them are 
too bright, so they are saturated in the VVV images. For the remaining two, 
the first variable, identified as source 23 in their work, has a mean $K$ 
magnitude of 11.25 and an amplitude of 2.95~mag, meaning that it should be 
detected in at least some of our images. In fact, it is present in our master 
catalog, but the number of good epochs is less than 25, probably because the 
magnitude of this star was close to the saturation limit. In consequence, it 
was rejected by our selection algorithm. The last one~-- source 90 in the 
\citetalias{woo98} paper~-- is an OH/IR star, identified as V4545~Sgr in the 
General Catalog of Variable Stars \cite[GCVS;][]{samu17}. The authors 
classify this star as a ``candidate LP'', and it is detected in our 
photometric catalog. The star is very red, with J-$K_S$ = 6.15; H-$K_S$=
2.11 and $K_S$= 11.03 (in 2010 images). 
Its amplitude is around 0.25 magnitude and was not considered as a variable object.  The follow up of this objecc (both photometry and  spectroscopy) is necessary in order to clarify its nature.
 

NV118 is a well-studied source, named V4524~Sgr in the GCVS. This object was originally detected as an OH/IR star in the work of \citet{lin92}, and it is also present as source 852 of \citetalias{mkn09}. Nevertheless, there is a discrepancy in the period found. While \citetalias{woo98} estimate a period of 885 days, we obtain 937 days. NV231 is another known OH/IR source, identified as V4541~Sgr in the GCVS. The authors calculate a period of 418 days, compared to our period of 464 days. In both cases, the discrepancies might be due to the fewer data points they had available, especially for NV231.

Most of our literature counterparts are included in the \citetalias{mkn09} work. There are 312 \citetalias{mkn09} objects in our field. Of these, 54 have $mflag = 2$ in their work, which implies they were too bright ($K_S<9$) and saturated. Figure~\ref{fig-11} presents the distribution of amplitude versus mean $K_S$ magnitude of the remaining 258 \citetalias{mkn09} sources in our field, represented by black circles. The vertical dashed blue line marks the typical saturation limit in VVV images, while the horizontal line is set at an amplitude of 0.2 mag, which is our lower limit to consider a star as variable. This implies that \citetalias{mkn09} stars located in the upper-right quadrant delimited by these lines should be detected by our method. Red circles indicate the matched variable stars of the present work. 

\begin{figure}
\centering
\includegraphics[width=\columnwidth]{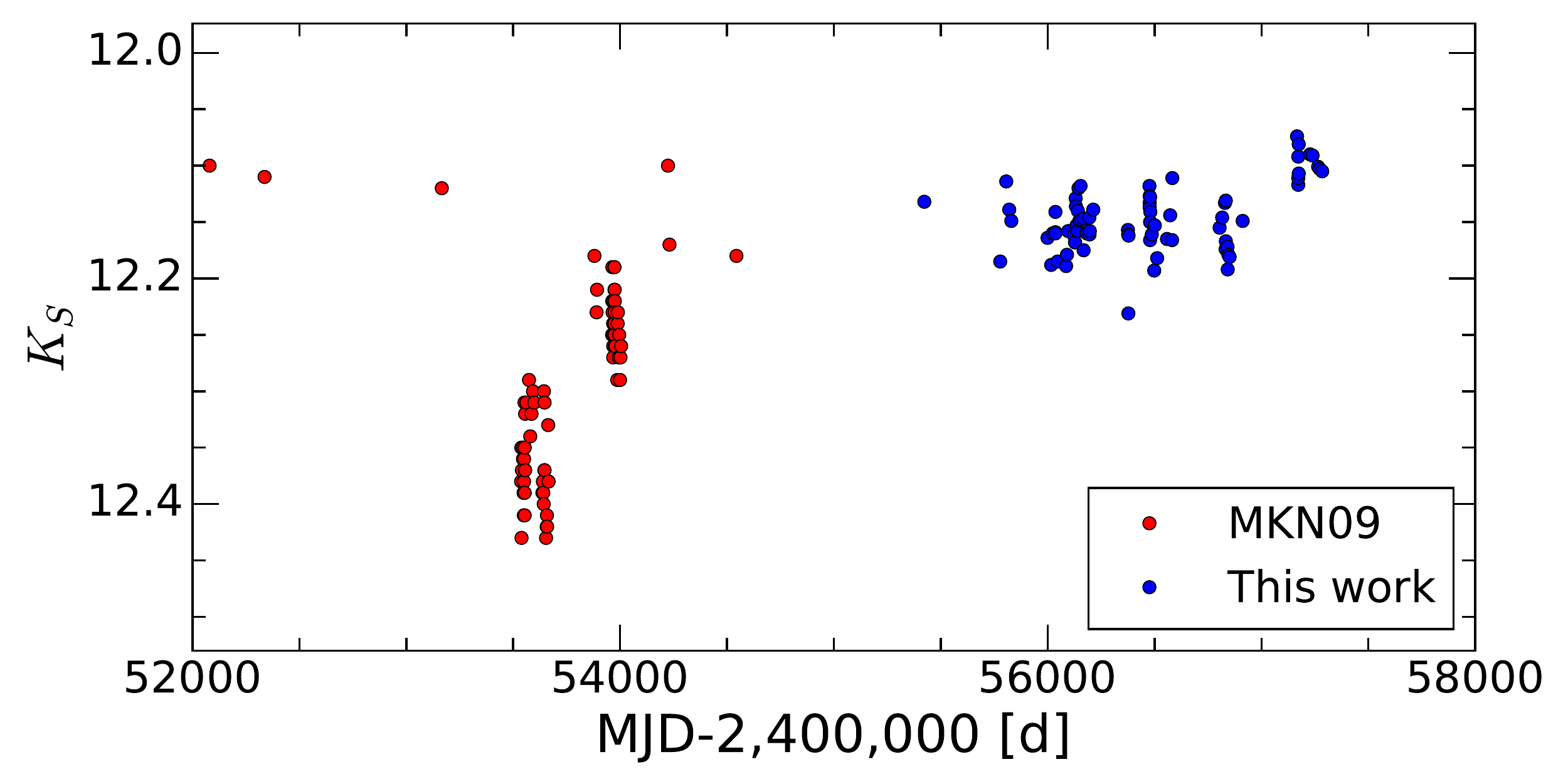}
\caption{Example of a source from the \citetalias{mkn09} catalog that was not classified as a variable star in the present work. Red circles correspond to \citetalias{mkn09} data, while blue circles represent VVV data.
\label{fig-12}}
\end{figure}

A cross-match with these variables shows that the majority of the \citetalias{mkn09} sources with ${K}_S<12.0$~mag do not even appear in our master photometric catalog. Since we required a reference epoch to assign coordinates in order to build the master photometric catalog, if a star was saturated in that epoch, it would not be considered for the remaining analysis. Hence, all these missing sources were saturated in the reference epoch. For the rest of the variable stars of \citetalias{mkn09} that are not present in our variability catalog, we did a one-by-one comparison. It turned out that almost all of these sources were included in the master catalog, however our selection method did not classify them as variable stars, either because the light curve had less than 25 epochs with good measurements~-- which was the case for stars with a mean magnitude close to the saturation limit~-- or because the amplitude was smaller than 0.2 mag. There are some remarkable cases where a star could have a large variability in the \citetalias{mkn09} light curve, while in our photometry the amplitude was too small. This is illustrated in Figure~\ref{fig-12}. The light curve shows a star with mostly small-amplitude variations, but two prominent drops in 2005 and 2006. The peak-to-peak $K_S$ amplitude of this star in the \citetalias{mkn09} catalog is 0.33~mag, while the amplitude in our master catalog is only 0.16~mag, rendering it below the amplitude cut explained in Section~\ref{sec:candsel}.

On the other hand, we have NV247, a star that has $K_{S,\text{MKN09}}=10.77$~mag, but from our light curve we see an important decrease in its brightness that makes it visible in our images. The final number of common candidates with \citetalias{mkn09} was 49. Of these, the authors obtained periods for only two sources, while we were able to calculate reliable periods for 14 variable stars with counterparts in their study.

\begin{table*}
\begin{center}
\caption{Feature comparison of \citetalias{mfk13} counterparts. See text for columns description.}
\label{tbl-2}
\begin{tabular}{cccccccc}
\hline
Name & No.$_\text{MFK13}$ & $K_S$ & $K_{S,\text{MFK13}}$ & $P$ & $P_\text{MFK13}$ & $P_\text{comb.}$ & Class \\
 & & (mag) & (mag) & (d) & (d) & (d) & \\
\hline
NV036 & 21 & 13.16 & 13.31 & 0.55646 & 0.55648 & 0.55647 & EB \\
NV142 & 28 & 12.75 & 12.95 & 15.567 & 15.543 & 15.556 & Cep(II) \\
NV150 & 29 & 13.31 & 13.58 & 10.268 & 10.260 & 10.264 & Cep(II) \\
NV307 & 38 & 12.24 & 12.24 & 1.6484 & 1.6486 & 1.6485 & EB \\
\hline
\end{tabular}
\end{center}
\end{table*}

\begin{figure}
\centering
\includegraphics[width=\columnwidth]{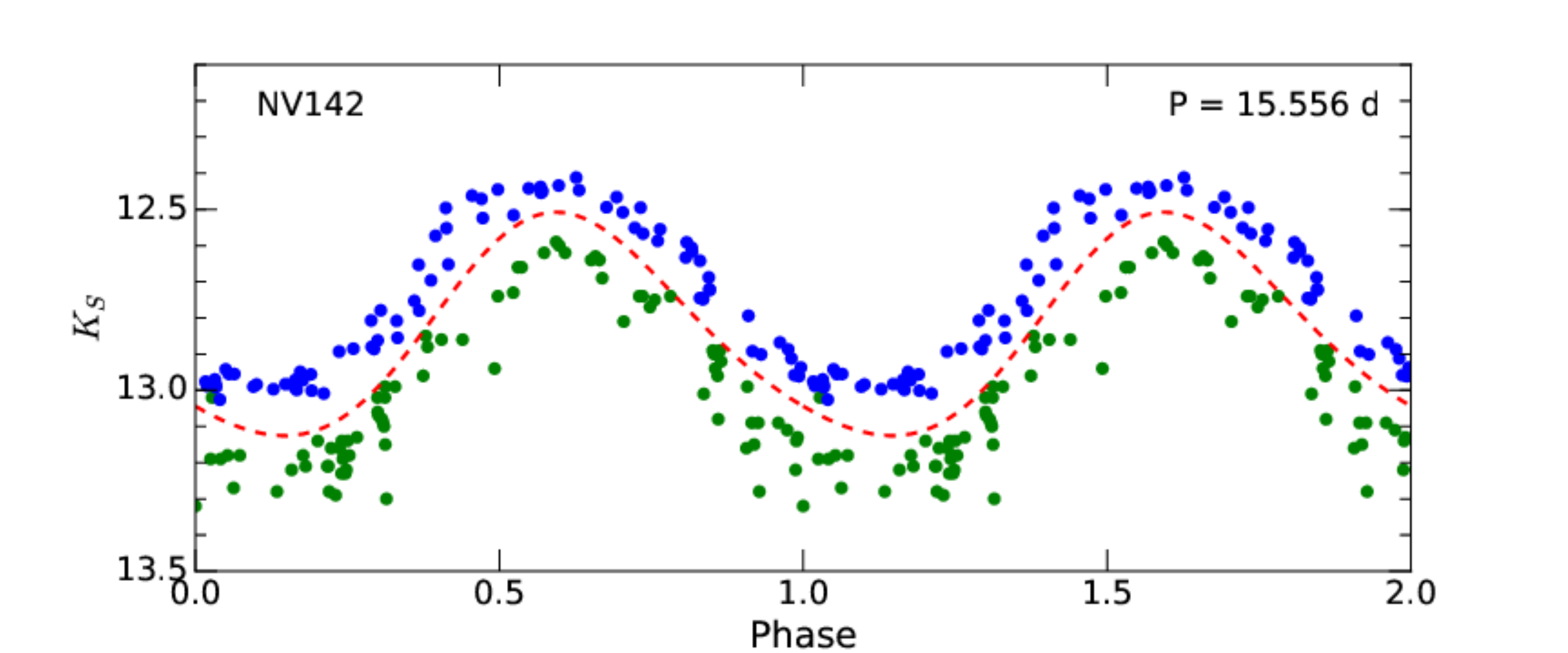}
\caption{Combined light curve of NV142 (blue circles) with source No.28 of \citetalias{mfk13} (green circles). The curve is folded with the period found using both data sets together. The red dashed line corresponds to a second-order Fourier fit.\label{fig-13}}
\end{figure}

The last work used for comparison is the research of \citetalias{mfk13}. The authors looked for short-period variables and Cepheids in the same area surveyed by \citetalias{mkn09}. We have 4 counterparts with their work: NV036, NV142, NV150 and NV307. The first and last are classified as EBs, while the other two are type II Cepheids. We checked the consistency of the calculated periods, and found a remarkable agreement; periods are similar to within 0.1\% of one other. In addition, we calculated periods using the combined light curves. These results are summarized in Table~\ref{tbl-2}. The first two columns show the variable names in each work. Columns 3 and 4 compare the mean $K_S$ magnitudes, while the following three columns include the periods of this work, \citetalias{mfk13} and combined, respectively. Column 7 presents the variability class, according to \citetalias{mfk13}. 

An important issue raised by this table, and confirmed when plotting the combined light curves, is a $\sim0.2$ mag shift between both data sets for most counterparts, apart from NV307. We consider that this shift~-- also present in some \citetalias{mkn09} counterparts~-- may be due to the differences between the photometric systems adopted in both studies. The transformation between VISTA and 2\,MASS are well known (see for example Soto et al. 2013, as well as the CASU Webpage), but they include a $(J-K_S)$ colour term. Unfortunately, the majority of counterparts in our case lack $J$ magnitudes; hence, in most cases, this colour term cannot be calculated. Another reason for the discrepancy could be the presence of differential reddening in the vicinity of these variables.  Nevertheless, an analysis of the light curves of regular variables in common with \citetalias{mkn09} and \citetalias{mfk13} shows that the typical shift between the 2\,MASS and VISTA systems ranges between 0 and $\pm0.2$~mag. For example, Figure~\ref{fig-13} shows a combined light curve for NV142. It is clear that our data, represented by blue circles, is $\sim0.2$ brighter than \citetalias{mfk13} data (green circles). The period used for the phased curve corresponds to the value of column 7 in Table~\ref{tbl-2}.

There is one additional \citetalias{mfk13} source that is located in our field, but is not present in our variable catalog. This object, listed as No. 23 in their paper, is detected by our photometry, though it has less than 25 good epochs.

In summary, there are 54 literature counterparts in our catalog. This means that we have 299 previously unknown variable stars, corresponding to 85\% of the total sample. Since \citetalias{mkn09} and \citetalias{mfk13} have public photometric data for their variables, we built combined light curves for all these stars.

Finally, it is worth remarking the advantages of combined light curves with the works of \citetalias{mkn09} and \citetalias{mfk13}. In addition to enabling a check of the derived periods, this also allows one to study the long-term variation of these stars, which is an issue that will be addressed later in the discussion.


\section{Red giant variables}
\label{sec:lpv}

\begin{figure}
\centering
\includegraphics[width=\columnwidth]{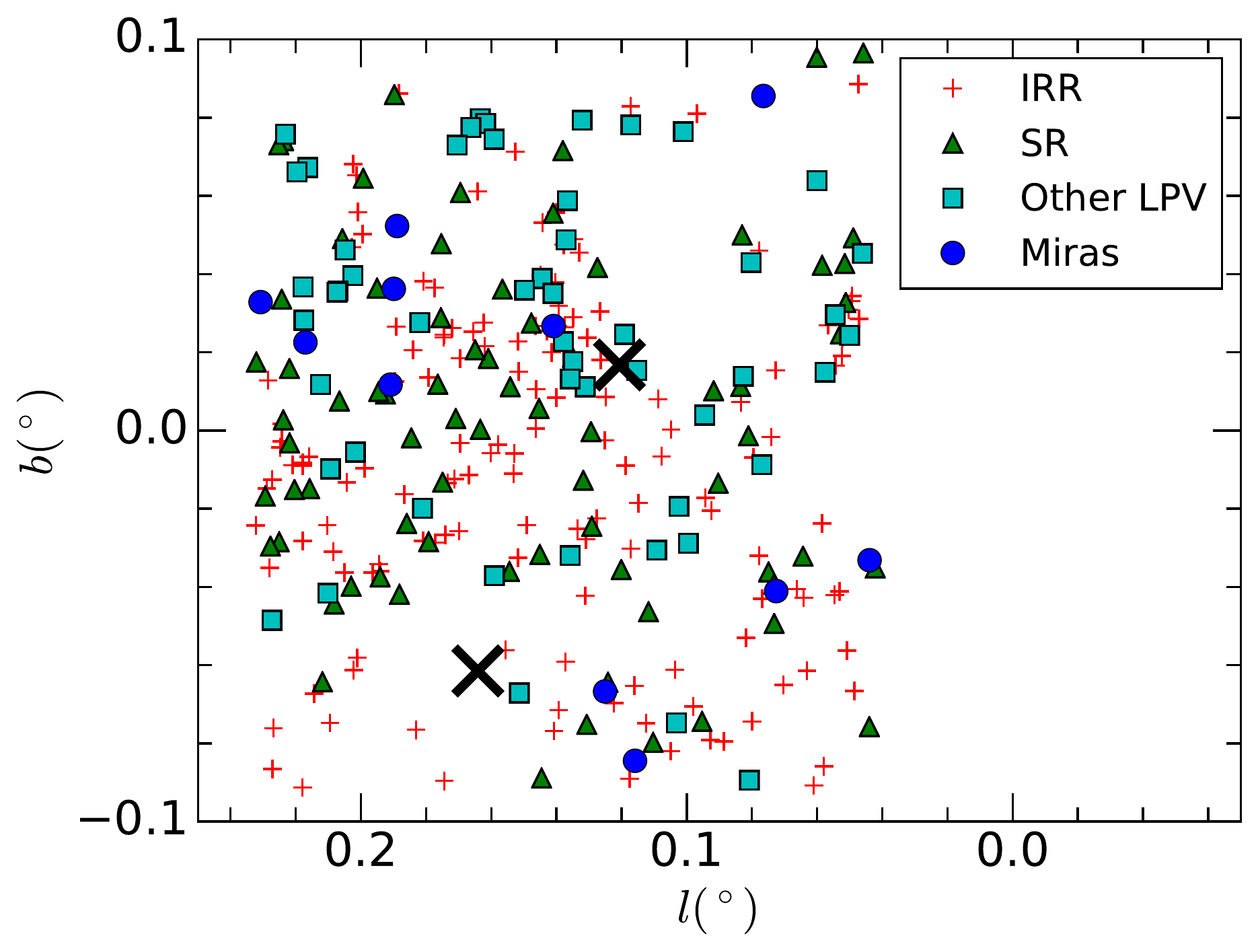}
\caption{Distribution of LPVs in the Galactic plane. Miras are represented as blue circles, SR correspond to green triangles, IRR variables are shown as red plus signs, and other LPVs are drawn with cyan squares. The two ``X'' symbols mark the position of Arches and Quintuplet, with Arches being closer to the image centre.\label{fig-14}}
\end{figure}

We already mentioned that there is a great diversity of variable classes expected in the near-IR. In the following sections we will discuss the main types found in this work. Variable stars were mainly classified according to light-curve morphology, though position in the CMD and colour-colour diagrams were also considered in some cases. We start our discussion with the most numerous class present in our catalog.

There is a total of 285 red variable stars, equivalent to 81\% of all variable stars in our catalogue. Of these, we have 11 Miras, 75 SR and 147 IRR variables. In addition, within our catalog we have found 52 stars that show a long-term increase or decrease in brightness. These objects are too bright to be YSOs belonging to the GC, and the probability that all of them are projected YSOs is low. Moreover, their light curves do not show clear features of pulsation, or rather the pulsation is not the dominant physical process behind their variability. Since their colours and magnitudes are consistent with red giant stars, we will classify them as ``other LPVs'', to enforce the concept that these are red giant stars, but they cannot unambiguously be put into the three categories already described, namely, Miras, SR or IRR variables. See Section~\ref{sec:trend} for a discussion about these variable stars.

It is important to note that the separation among these types is sometimes very fine, and this might lead to some incorrect classification. While Miras are easily distinguished from the other classes thanks to their large amplitude and nearly sinusoidal variation, SR and IRR stars are harder to tell apart. To discriminate between the latter, we used the GLS periodogram, as explained in Section~\ref{sec:period}. This means that a source that had a peak in the power spectrum smaller than 0.6 was classified as non-periodic, but this is an arbitrary limit. In addition, the sampling of the light curve heavily impacts the ability to perform a correct classification, especially when, due to the intrinsic non-uniform time spacing of the VVV images, there is missing information on maxima and/or minima. Of course, this is more relevant for LPVs than for shorter-period variables.

The location of these stars in the Galactic plane is shown in Figure~\ref{fig-14}. They follow a spatial distribution similar to what is displayed in Figure~\ref{fig-4}. Miras are mostly located towards the northeast section of the field.

\begin{figure}
\centering
\includegraphics[width=\columnwidth]{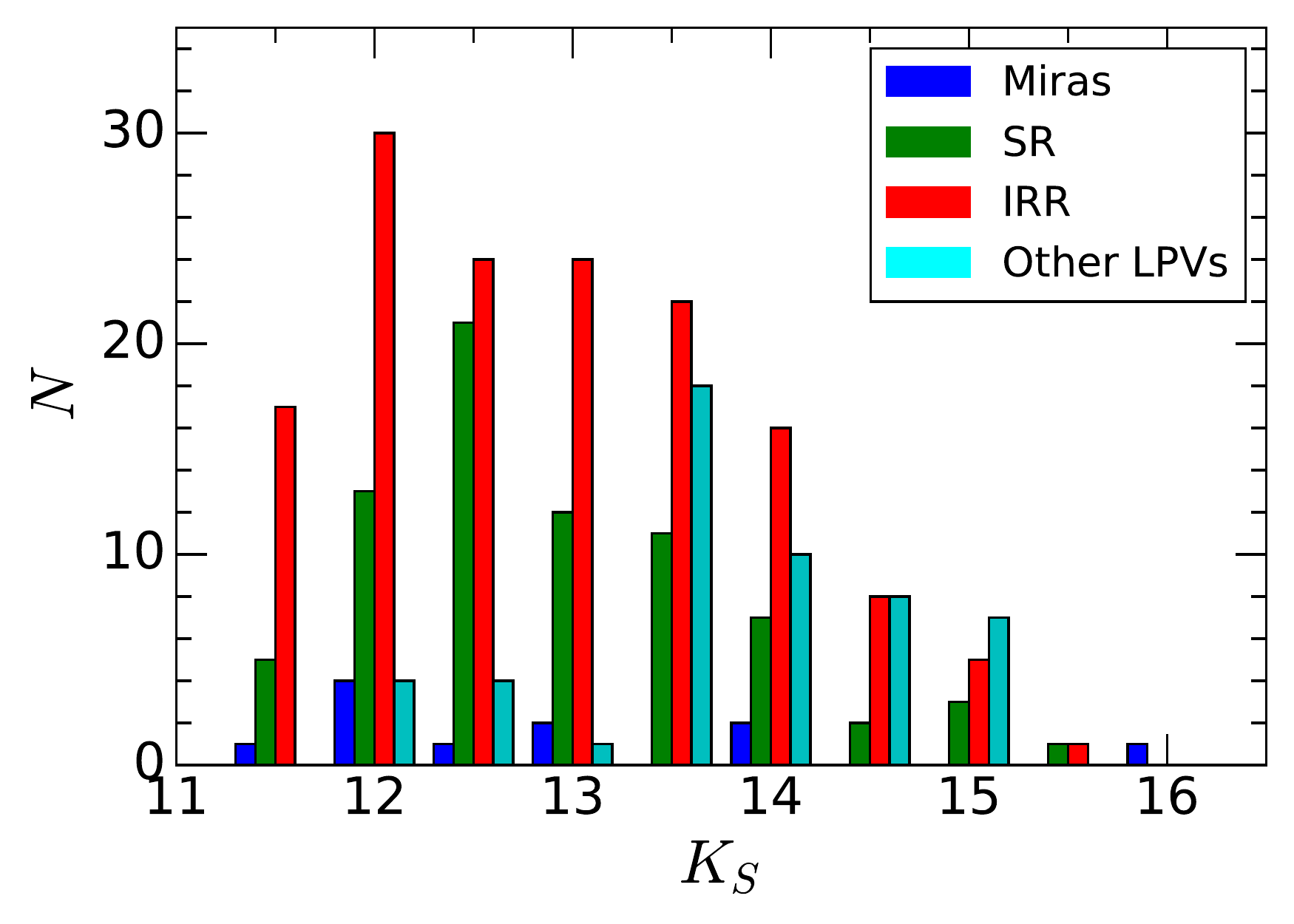}
\caption{Histogram of magnitudes for LPVs. The blue, green, red and cyan bars correspond to Miras, SR, IRR and other LPVs, respectively.
\label{fig-15}}
\end{figure}

Figure~\ref{fig-15} depicts the histogram for mean magnitudes of LPVs. The majority of red variables have magnitudes in the range of $11.5<K_S<14$. Mira variables with $\log(P)<2.6$ have mean absolute magnitudes between $-8<M_K<-6$ \citep*{whit08}. If we assume a distance to the GC of $\mu_0=14.52$~mag \citep[8~kpc,][]{ghez08}, and an average extinction of $A_{K_S}=2.5$~mag, we can estimate the apparent magnitude range of such variables to be between $K_S=10-12$~mag. This is without taking into account circumstellar extinction, which can move Miras towards fainter magnitudes.

It has been known that LPVs fall within well-defined sequences in a period-luminosity diagram \citep[e.g.,][]{wood99,sosz13}. In these so-called ``Wood diagrams'', Mira stars fall in sequence C, while SRs follow C and C' lines. On the other hand, SR variables have similar magnitudes to Miras \citep[e.g.,][]{sosz09}, with the exception of SRc stars, that are more luminous than Miras \citep{cat15}. In conclusion, the magnitude interval observed in Figure~\ref{fig-15} is consistent with the majority of these AGB stars belonging to the GC. 

Miras distribute uniformly across magnitudes, though their number is too small to make further analysis. On the other hand, SRs have a peak at $K_S=12.5$ mag. Towards fainter magnitudes the number of SR stars decreases, while the number of IRR variables remains more stable. This is likely caused by fainter stars being harder to classify as periodic (or semi-periodic) due to the presence of a larger scatter in magnitude. It is interesting that other LPVs have a strong peak at $K_S=13.5$ mag, while they have small numbers in the remaining bins. 


\subsection{Atmospheric abundances of AGB stars}

\begin{figure}
\centering
\includegraphics[width=\columnwidth]{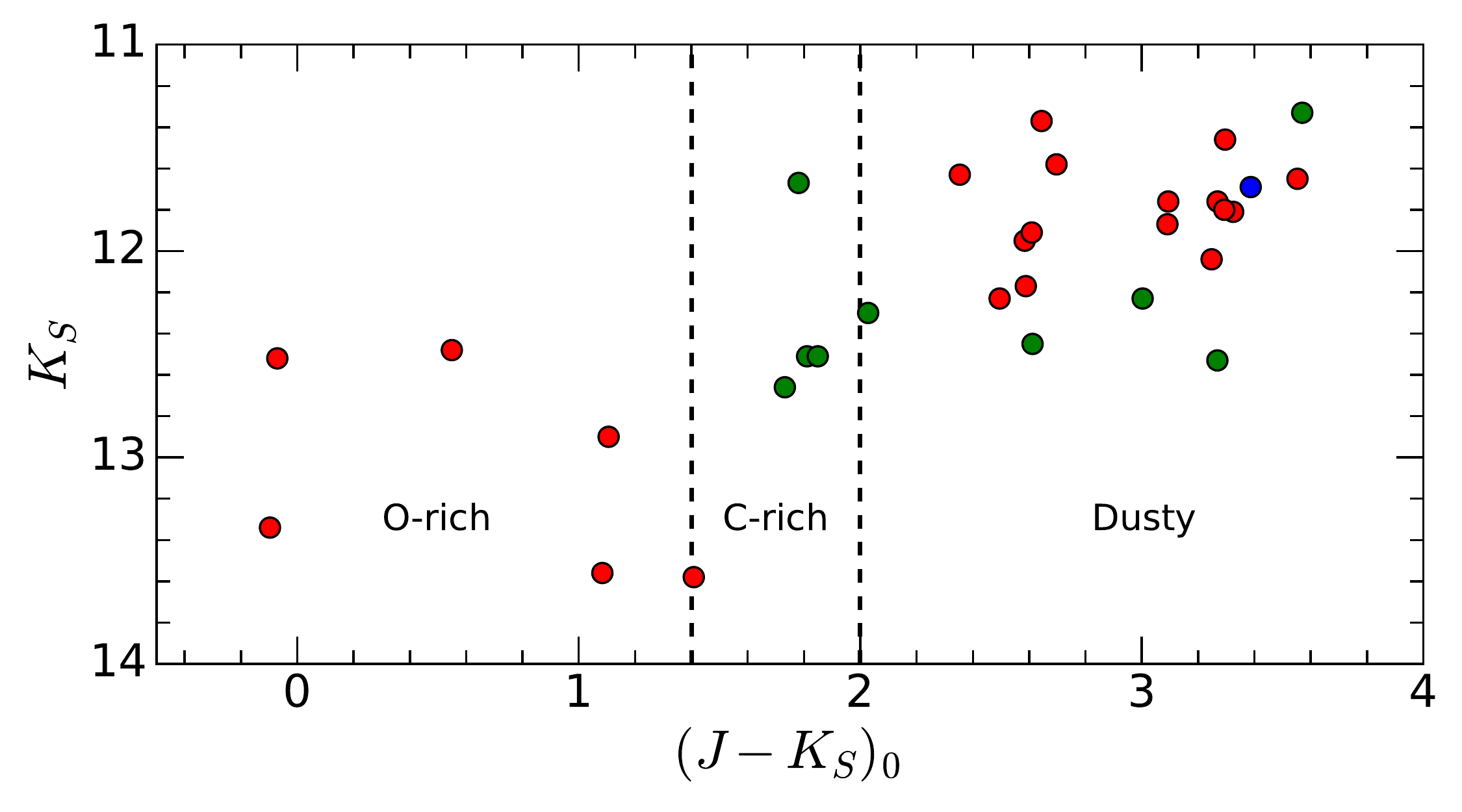}
\caption{$(J-K_S)_0-K_S$ diagram for AGB stars. colours correspond to the variability type: Miras (blue), SR (green) and IRR (red). Vertical lines at $(J-K_S)_0=1.4$~mag and $(J-K_S)_0=2$~mag mark the separation between O-rich, C-rich and dusty AGB stars (see text for details).
\label{fig-16}}
\end{figure}

Without a spectroscopic follow-up, another method to separate between C-rich and O-rich AGB stars is through their IR colours. For example, the mid-IR colour $[9]-[18]$ effectively distinguishes these two populations when stars are surrounded by thick dust envelopes \citep[e.g.,][]{ita10,mats17}. While carbon stars present SiC emission at 11.3\,\micron\ plus a continuum excess produced by amorphous carbon, O-rich stars show silicate bands at 9.8 and 18\,\micron\ \citep{mats17}. Since these bands are not available for the variables in this study, an alternative is the use of the {\it Spitzer} IRAC bands, as shown in Section~\ref{sec:mid-ir}.

In the near-IR, a common criterion is the use of the $(J-K_S)$ colour \citep[e.g.,][]{sosz05}. The differences between C-rich and O-rich stars in these filters is that the former have strong C$_2$ and CN bands present in $J$, with no relevant bands in $K_S$. On the other hand, the H$_2$O absorption is strong both in the $J$ and $K_S$ filters, added to the presence of TiO and VO bands in $J$. The usual criterion is to consider stars with $(J-K_S)_0<1.4$ as O-rich; if $1.4<(J-K_S)_0<2$, they are classified as C-rich. Stars with redder colours are surrounded by thick shells and can belong to any of these two types \citep[][and references therein]{sosz05}. Colours have to be dereddened to properly apply the previous criterion. To do so, we used the extinction map provided by \citet{gonz12} to estimate the colour excess for each of the 31 AGB stars with $J$ magnitudes. The dereddened diagram is illustrated in Figure~\ref{fig-16}. Stars with $(J-K_S)<0$ are probably foreground RGB stars with lower extinction, and therefore will be excluded from this analysis. NV061, NV091 and NV218 are IRR variables and have colours consistent with O-rich AGB stars. NV206 is almost at the separation between both atmospheric chemistry, so it cannot be classified properly. In this regard we must note that the near-IR colours presented here are calculated with a single measurement in $J$, so this introduces some scatter to the plot. If we add this to the typical photometric errors, classification of sources is uncertain within $\sim0.1$ mag from the colour-cut lines.

There are four SR variables with colours of C-rich stars, namely, NV033, NV056, NV239 and NV247. The case of NV033 is worth discussing, since its period and amplitude are consistent with Mira variables, though the light curve is not completely regular. However, there is evidence of cycle-to-cycle variations in C-rich Miras \citep{whit03}, hence the possibility of a carbon Mira cannot be ruled out. In any case, the period and magnitude of this star would put it behind the GC. These four stars are added to the six sources discussed in Section~\ref{sec:mid-ir} with mid-IR colours consistent with C-rich stars.

The rest of the sources plotted in Figure~\ref{fig-16} are dusty AGB stars, so it is not possible to derive their atmospheric C/O ratio with near-IR colours.


\subsection{Miras and semi-regular variables}

\begin{figure}
\centering
\includegraphics[width=\columnwidth]{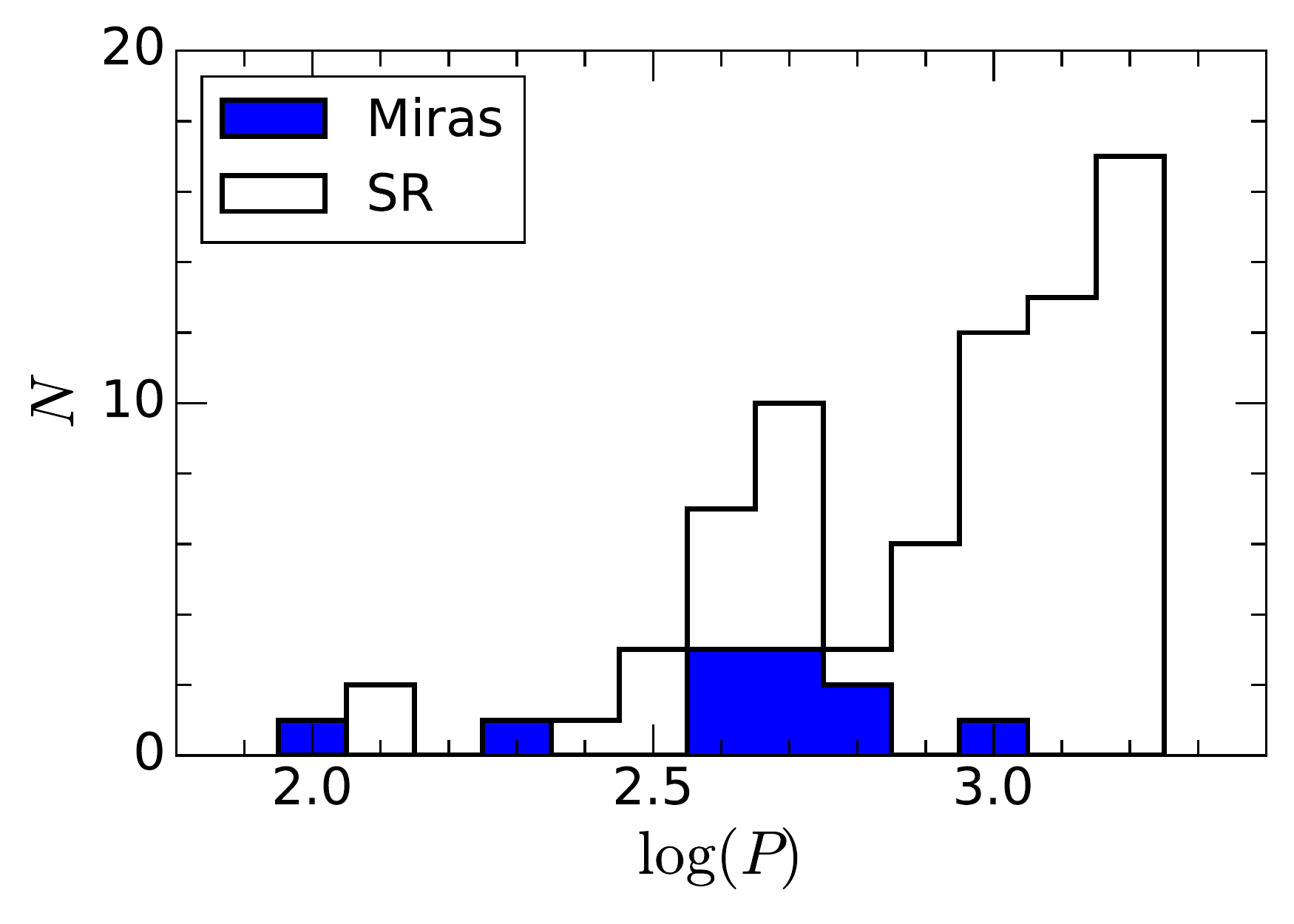}
\caption{Period distribution for LPVs. The period histogram for Miras is shown in blue, while the histogram for SR variables is shown in white.\label{fig-17}}
\end{figure}

The period distribution for Miras and SR stars is presented in Figure~\ref{fig-17}. Mira variables have a peak at $\log(P)=2.7$, while no such stars are detected between $2.4<\log(P)<2.6$. In comparison, Miras from \citetalias{mkn09} peak at $\log(P)\sim2.5$. A similar value was found by \citet{catch16} for more than 8000 Miras in the Galactic bulge. Nevertheless, this is an expected outcome, despite the small number statistics. If we assume all these Miras are located at the distance of the GC or nearby, their magnitude span should be the one mentioned a couple of paragraphs above. At $\log(P)\sim2.3$, $K_S\sim11$~mag, meaning that these stars are saturated in the VVV images, a situation already discussed in \S~\ref{sec:literature} when comparing with the \citetalias{mkn09} counterparts. The presence in our catalogue of Miras with periods between 200 and 400 d would immediately suggest that those stars should be located behind the GC.

Regular variables present in this catalog with $\log(P)>2.6$ will be called ``infrared Miras'' for the remainder of this analysis. These are AGB stars with thick dust shells, which produces a very large circumstellar extinction, even in near-IR bands. In consequence, they do not have optical counterparts. The dimming is strong enough to put these objects below the $K_S$-band saturation limit of the VVV images. In our Galaxy, these objects are mostly related with OH/IR stars.

Infrared Miras do not follow the same PL relation, unless they are corrected for circumstellar extinction \citep{ita11}. Miras with $\log(P)>2.6$ deviate significantly from the period-luminosity line, a separation that is even stronger for C-rich Miras. This effect is less relevant in mid-IR passbands; however, for the $K_S$-band the deviations are important enough to be treated with care when estimating distances. \citet{ita11} suggest that the strong circumstellar extinction may affect the PL relation. As a consequence, any deviation from the predictions should depend on the circumstellar extinction. The authors relate this deviation with near-IR colours, so a correction can be applied to distance estimations. 

Semiregular variables have two peaks in Figure~\ref{fig-17}, one at $\log(P)=2.6$ and another, more prominent, at $\log(P)=3.2$. The presence of SR variables with $K_S>15$~mag indicates that such sources are beyond the GC.

There are 11 Mira variables present in our catalog. In addition, there are two more Miras discussed in \citetalias{nav16}, namely, VC17 and VC19. Three out of these 13 sources do not have literature counterparts, VC17, NV103 and NV226. NV231 is present in the \citetalias{woo98} catalog, while the others were observed by \citetalias{mkn09}. However, only three of them have period estimations in those papers. On the other hand, NV118, NV166, NV231 and NV295 are classified in the literature as OH/IR~-- hence O-rich~-- stars. 


\subsection{Period-amplitude relation}

\begin{figure}
\centering
\includegraphics[width=\columnwidth]{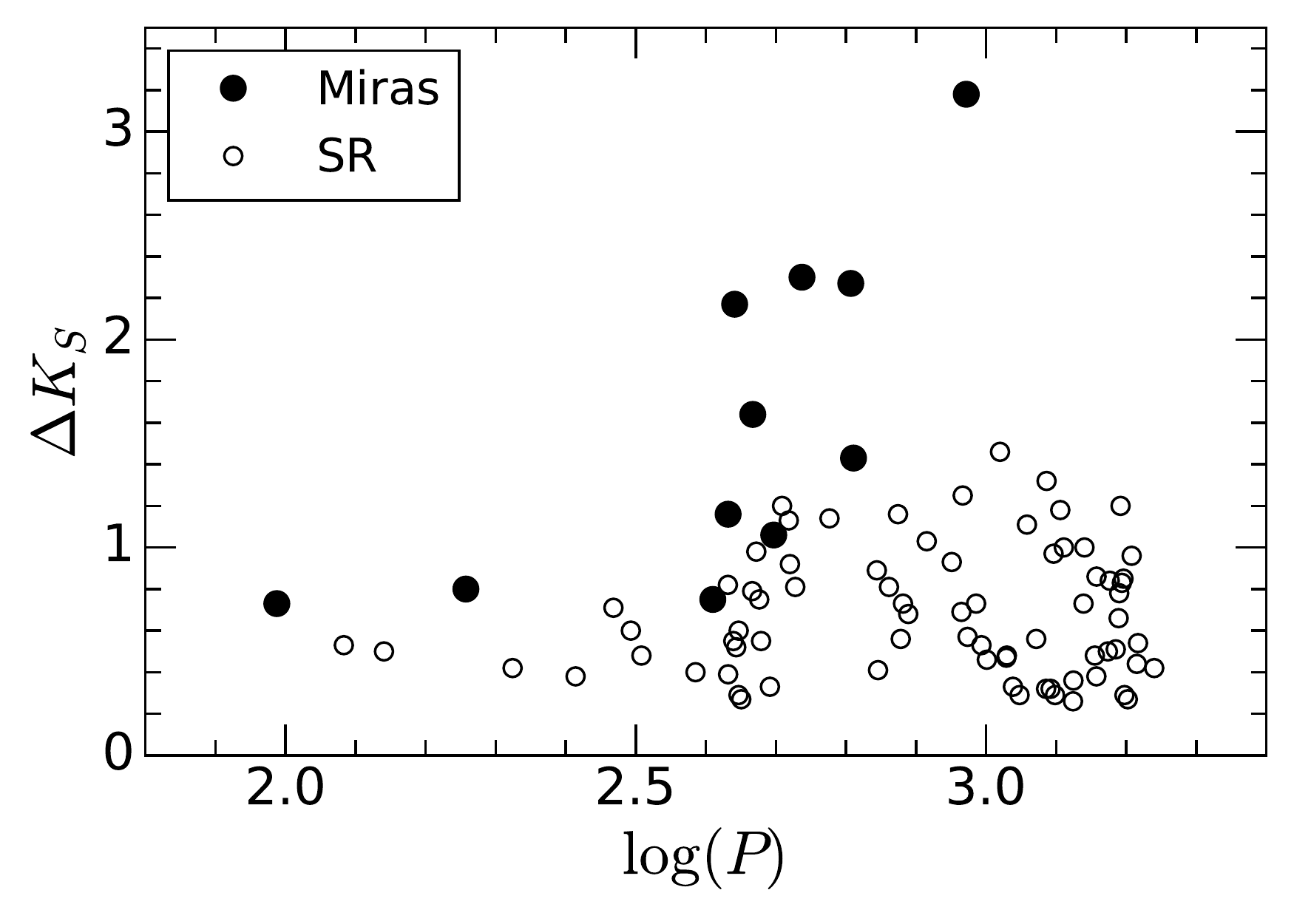}
\caption{Pulsation amplitude versus period (in days) for LPVs. Black circles correspond to Miras, while open circles are SR variables.\label{fig-18}}
\end{figure}

\begin{figure}
\centering
\includegraphics[width=\columnwidth]{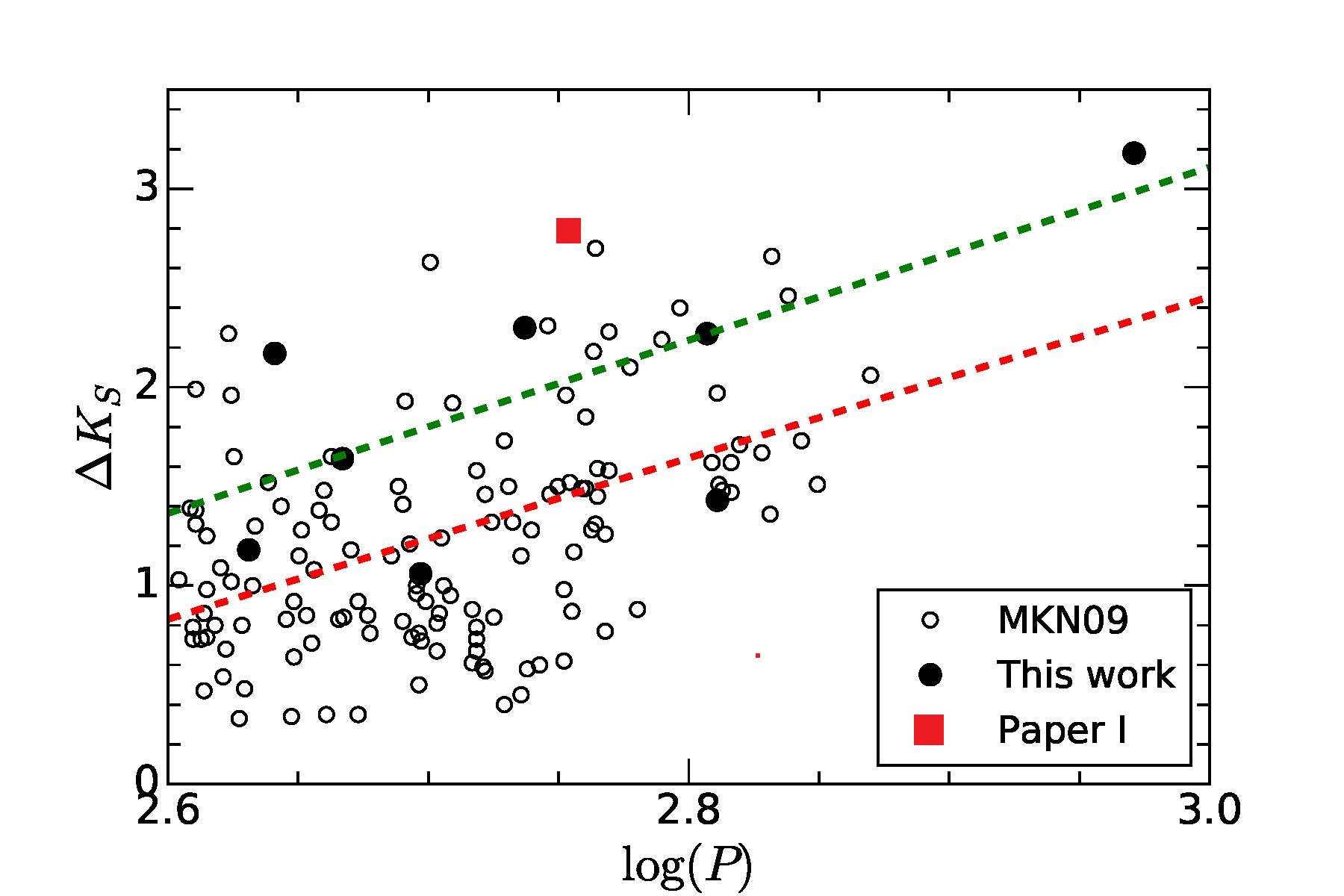}
\caption{Comparison of amplitudes and periods for our Miras (full circles) with $\log(P)>2.6$ and those from \citetalias{mkn09} (open circles). The red dashed line corresponds to a linear fit for the combined data, while the green line is a linear fit for the Miras found in our study, plus VC17 from \citetalias{nav16}. The latter star is plotted as a rectangle.\label{fig-19}}
\end{figure}

The relation between peak-to-peak amplitude and period is plotted in Figure~\ref{fig-18}. SRVs (open circles) present a large scatter in the diagram. As expected, these variables have smaller amplitudes than Miras for a given period. Miras (full circles) show a correlation of increasing amplitude with period, which is stronger for $\log(P)>2.6$. This trend has been observed in other Mira surveys \citep[see e.g., Fig. 3 of][]{whit03}.

In order to perform a further analysis of this correlation, we included all Mira variables from the \citetalias{mkn09} catalog with $\log(P)>2.6$. The results are illustrated in Figure~\ref{fig-19}. The green line corresponds to a linear fit performed over the 8 Miras with $\log(P)>2.6$ described in this work, with the addition of VC17, which was presented in~\citetalias{nav16}. The expressions for the linear fits represented by the green and red lines are
\begin{equation}
\label{eq-per_amp1}
  \Delta K_S = (4.36 \pm 2.33) \log(P) - (9.97 \pm 6.39) \,,
\end{equation}
\begin{equation}
\label{eq-per_amp2}
  \Delta K_S = (4.06 \pm 0.64) \log(P) - (9.73 \pm 1.73) \,,
\end{equation}
respectively. The regressions assume that the period is the independent variable. Even though the amplitude scatter in the diagram is large, there is a clear tendency that is shared between both data sets. On average our variables have amplitudes $\sim0.6$~mag larger than those from \citetalias{mkn09}; however, they are still within the range observed in the \citetalias{mkn09} catalog. We must note that this relation is poorly constrained in the long-period end, since there are two variables with $\log(P)>2.85$, and only one star with $\log(P)>2.9$, which is NV118. The characterization of more Miras with very long periods would greatly improve the accuracy of Eqs.~\ref{eq-per_amp1} and~\ref{eq-per_amp2}.

The large spread in amplitudes is probably caused, at least in part, by the use of peak-to-peak amplitudes. Since we are in the infrared Miras regime, dust obscuration events will produce significant changes in the mean magnitude from one cycle to another, leading to larger amplitudes \citep[e.g.,][]{whit03}. A more consistent approach would be to use the mean amplitude per cycle instead, calculated through a Fourier fit.

In addition, \citetalias{mkn09} do not distinguish between Mira and SR variables, so they are all treated as Mira stars, regardless of the features shown in their light curves. This is corroborated through the visual inspection of the combined light curves, where their stars present features of Miras, SR and IRR variables. Since SR variables have, on average, smaller amplitudes than Miras \citep*[e.g.,][]{sosz13b}, this will contribute to the wide spread in amplitudes observed in Figure~\ref{fig-19}.


\section{Distances and extinctions}
\label{sec:dist}

\begin{figure}
\centering
\includegraphics[width=\columnwidth]{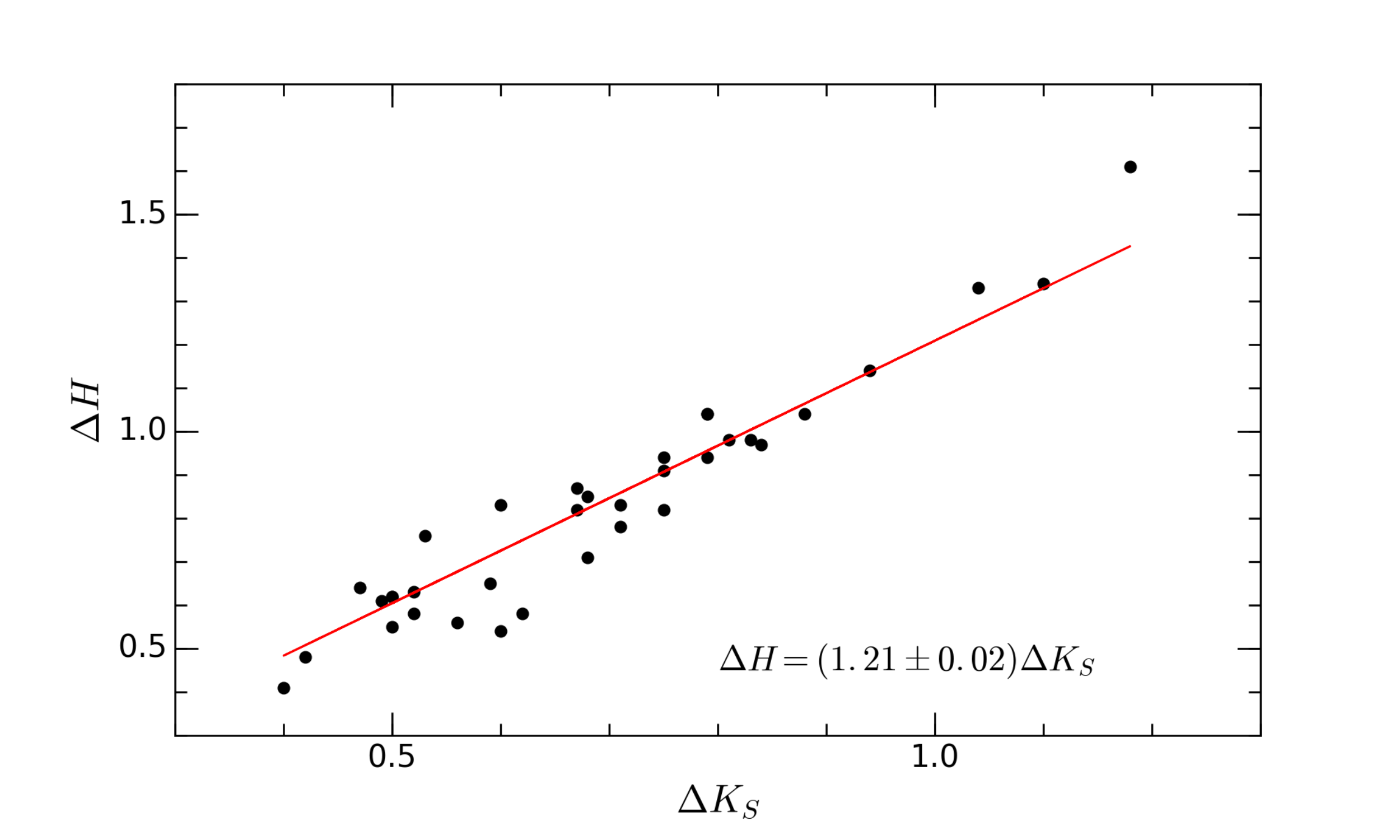}
\caption{$H$ versus $K_S$ amplitudes for periodic variables of the \citetalias{mkn09} catalog. The red line is a linear fit to the data.\label{fig-20}}
\end{figure}

\begin{table*}
\centering
\caption{Distances and extinctions for Mira variables studied in this work, along with one Mira from \citetalias{nav16}.}
\label{tbl-3}
\small
\begin{tabular}{lccccccccccc}
\hline
Name & $P$ & $M_{K_s}$ & $A_{K_S,\text{NNK06}}$ & $\mu_{0,\text{NNK06}}$ & $D_{\text{NNK06}}$ & $A_{K_S,\text{NTH09}}$ & $\mu_{0,\text{NTH09}}$ & $D_{\text{NTH09}}$ & $A_{K_S,\text{AMC17}}$ & $\mu_{0,\text{AMC17}}$ & $D_{\text{AMC17}}$\\
 & (d) & (mag) & (mag) & (mag) & (kpc) & (mag) & (mag) & (kpc) & (mag) & (mag) & (kpc) \\
\hline

NV005 & 498 & -8.29 & 5.74 & 14.55 & 8.14 & 6.74 & 13.55 & 5.13 & 4.82 & 15.46 & 12.4 \\
NV106 & 438 & -8.09 & 4.07 & 16.82 & 23.1 & 4.79 & 16.11 & 16.7 & 3.43 & 17.47 & 31.2 \\
NV166 & 546 & -8.43 & 4.51 & 16.80 & 23.0 & 5.30 & 16.01 & 15.9 & 3.79 & 17.52 & 32.0 \\
NV188 & 407 & -7.98 & 3.77 & 16.36 & 18.7 & 4.42 & 15.70 & 13.8 & 3.17 & 16.96 & 24.6 \\
NV222 & 97 & -5.76 & 3.05 & 14.40 & 7.60 & 3.58 & 13.87 & 5.94 & 2.56 & 14.89 & 9.50 \\
NV226 & 428 & -8.06 & 3.82 & 16.15 & 17.0 & 4.49 & 15.48 & 12.5 & 3.21 & 16.76 & 22.5 \\
NV231 & 465 & -8.18 & 2.81 & 19.50 & 79.7 & 3.30 & 19.01 & 63.6 & 2.36 & 19.95 & 98.0 \\
NV315 & 181 & -6.73 & 5.21 & 13.69 & 5.48 & 6.12 & 12.78 & 3.60 & 4.38 & 14.52 & 8.03 \\
VC19 & 111 & -5.97 & 3.83 & 14.23 & 7.01 & 4.50 & 13.56 & 5.15 & 3.22 & 14.84 & 9.29 \\
\hline
\end{tabular}
\end{table*}

With the aim of measuring accurate distances to Mira variables, we have to derive the extinction towards these stars. We will use the method carried out in \citetalias{mkn09} to calculate both extinctions and distances. For this, we require $H$ and $K_S$ light curves. The VVV survey has only two epochs of observations in the $H$-band. However, the mean value of $H$ across the complete pulsation cycle can be extrapolated by assuming that light curves of Miras in the near-IR differ exclusively in their amplitudes, and not in their shape or phase. To corroborate this assumption, we compared the $J$, $H$ and $K_S$ light curves of periodic variables from the \citetalias{mkn09} catalog, since their observations are simultaneous. In general, the light curve shapes in these bands do not differ significantly.

After this, we derived a relation between the $H$ and $K_S$ amplitudes for this group of periodic variables, using data from \citetalias{mkn09}. The results are illustrated in Figure~\ref{fig-20}. The red line corresponds to a linear fit to the data, with the condition that the fit passes through the origin. The resulting equation is
\begin{equation}
  \Delta H = (1.21 \pm 0.02) \Delta K_S \,.
\end{equation}
In conclusion, we can say that Miras have $H$ amplitudes $\sim20\%$ larger than in the $K_S$-band. With this relation, combined with the pulsation period and the single-epoch $H$ magnitude, we are able to infer a mean magnitude value in $H$.

For the purposes of this work, we will use the PL relations shown in Equations 2 and 3 of \citetalias{mkn09}. With these two equations, the expressions for the extinction $A_{K_S}$ and the true distance modulus $\mu_0$ correspond to Equations 8 and 9 of \citetalias{mkn09}.

Before we derive the distances to our Mira variables, it is relevant to discuss the effect of the extinction law  on the results. \citet{riek85} derived the first extinction law towards the GC. The authors found a total-to-selective extinction ratio $R_V=3.09$, very similar to the ``universal'' value of 3.1. Later on, \citet{udal03} used red clump (RC) bulge stars found on OGLE II images to derive a new extinction law. The value of $R_V$ turned out to be surprisingly low and variable: between 1.75 and 2.5. 

A new extinction law in the near-IR was derived by \citet[][hereafter NNK06]{nish06} using the RC method. This work reinforced the notion that extinction towards the GC varies among different lines-of-sight, at least in the IR regime. Later on, \citetalias{nish09} used 2MASS data of RC stars to improve the results of \citetalias{nish06}, and extended the extinction law to the mid-IR bands of {\it Spitzer}. Finally, \citet[][hereafter AMC17]{alon17} derived new near-IR extinction ratios for the inner Galaxy, using data from the VVV survey. In this case, the authors found that the wavelength dependence of the extinction followed a steeper power-law than the values reported by \citetalias{nish06} and \citetalias{nish09}.

Since there is no agreement about the correct extinction law towards the GC, we decided to derive our distances using the extinction ratios from the three studies mentioned in the previous paragraph, in order to compare the different results. If we define $r_H \equiv A_H / A_{K_S}$, we obtain $r_{H,\text{NNK06}}=1.74$, $r_{H,\text{NTH09}}=1.63$, and $r_{H,\text{AMC17}}=1.88$. For the sake of consistency, the first two values have been corrected for the VISTA photometric system.

Out of the 11 Mira variables reported in this work, we have $H$ magnitudes for 8 of them. In addition, we included the Mira VC19 from \citetalias{nav16}. Table~\ref{tbl-3} presents a summary of the extinctions and distance moduli for these 9 variable stars, using the three extinction laws mentioned above. Column 1 shows the variable star name assigned in the present work or in \citetalias{nav16}. Column 2 presents the period calculated with the GLS periodogram, and column 3 contains the mean absolute $K_S$ magnitude, from Eq.~3 of \citet{mkn09}. The extinction, distance modulus and distance using the NNK06 extinction law are shown in columns 4, 5 and 6. Columns 7, 8 and 9 contain the same values, but for the NTH09 law. Finally, the values calculated using the AMC17 extinction law are displayed in columns 10, 11 and 12.

The influence of the chosen extinction law over the distances obtained is clear. When the \citetalias{nish09} law is used, Miras are systematically closer, and all extinctions are larger. If we take a look at the two variables with shorter periods so the PL calculations are not greatly affected by circumstellar extinction, namely NV222 and VC19, we see that, if we use the \citetalias{nish06} or the \citetalias{alon17} law, they are located near the GC, as expected. However, when the \citetalias{nish09} law is utilized, they are seen as foreground objects located at $\sim5$~kpc from us. These findings suggest that the \citetalias{nish09} law is less suitable to model the extinction towards the GC, though more detailed studies will be required to conclusively solve this discrepancy.

The PL relations for Miras found in the literature are usually valid for $\log(P)\lesssim2.6$ \citep[e.g.,][]{whit08,mkn09}. We have discussed the effect of circumstellar extinction on longer-period Miras several times across this paper, in terms of the observed deviations from the PL relation. It is clear that all the infrared Miras from Table~\ref{tbl-3}, with the exception of NV005, have overestimated distances. As observed by \citet{ita11}, a thick dust shell would shift Miras below the PL line, and this effect would be more noticeable when the envelope is thicker. On the other hand, this relation is not as well established as, e.g., the one followed by Cepheids (e.g., Whitelock 2013). Thus, although in general we can expect that the presence of a dust shell will change the distance moduli for infrared Miras, more data are also necessary to better establish the Mira PL relation and its range of applicability. Recently, Mowlavi et al. (2018) published the first Gaia catalog of LPVs, and improved Gaia distances could be of help in this context. However, the fact that Miras are physically very large and can have inhomogeneous surfaces can induce large errors in the thus derived distances, if such inhomogeneities are not properly taken into account (Whitelock \& Feast 2014).

Finally, it is worth commenting the case of NV231, since it has the smallest extinction of all Miras. The distance calculated for this star puts it far behind the GC. If we assume that the circumstellar extinction of this star accounts for a difference of 2 mag in the PL relation, this would imply a corrected distance of 31.6~kpc, so this Mira variable would belong to the far side of the disk. This star seems to be located in a region with a small foreground extinction, which is finally the reason why we can observe it, in spite of its being further away from the GC.
  
\section{Comparison with Gaia DR2 distances}

The Gaia astrometric mission was launched in December 2013 \citep{2016A&A...595A...1G} to measure positions, parallaxes, proper motions and photometry for over $10^9$ sources as well as to obtain physical parameters and radial velocities for millions of stars.  
Its recent Data Release 2 (Gaia DR2) has covered the initial 22 months of data-taking \citep{gaia-collaboration2018}. The Gaia distances for 1.33 billion stars were recently published by Bailer-Jones et al. (2018). The cross-identification within 1 arcsec radius (selected to minimize the crowding) shows, as expected, only 2 stars in common (NV091 and NV294), of which only NV091 seems to belong to the GC region.
We do not have any variables in common with the Mowlavi et al. (2018) catalog of LPVs based on Gaia DR2.

\subsection{Error estimations}

The errors in the calculated distances and extinctions come from several different sources: the standard deviation (SD) of the PL relation \citepalias[Eqs.~2 and~3 from][]{mkn09}, the errors in the calculated coefficients of the reddening laws, uncertainties in the period estimations, and finally the photometric errors of the Mira variables.

The SDs of those equations are 0.19 and 0.17~mag, respectively \citepalias{mkn09}, and they represent the typical scatter of the PL relations. According to \citetalias{mkn09}, these SDs will introduce errors of 0.08 and 0.18 mag in $A_{K_S}$ and $\mu_0$, respectively.

Periods calculated have errors that depend on the period itself. For example, Miras with $P>400$~d have errors of $\sigma_{P}\sim20$~d. This translates into errors of 0.01~dex in $\log(P)$. In consequence, the added error to the PL equations is $\sim0.04$~mag. The period errors will barely affect the calculated extinction, since they are less than 0.01~mag. Additionally, their effect on the distance modulus will be $\sim0.04$~mag. 

On the other hand, the errors in the extinction coefficients for the \citetalias{nish06} and \citetalias{nish09} laws are about $1-2\%$ \citep{nish06,nish09}. This translates into systematic uncertainties of 0.06 and 0.07~mag for Eqs.~8 and 9 from \citetalias{mkn09}. Finally, photometric uncertainties for the magnitudes of Miras in this work range between 0.02 and 0.05~mag.

Adding these random and systematic uncertainties in quadrature gives the total errors for $A_{K_S}$ and $\mu_0$. The final values are $\sigma_{A_{K_S}}\approx0.11$~mag and $\sigma_{\mu_0}\approx0.20$~mag.


\section{Long-term trends in red giants}
\label{sec:trend}

\begin{figure*}
\centering
\includegraphics[width=\textwidth]{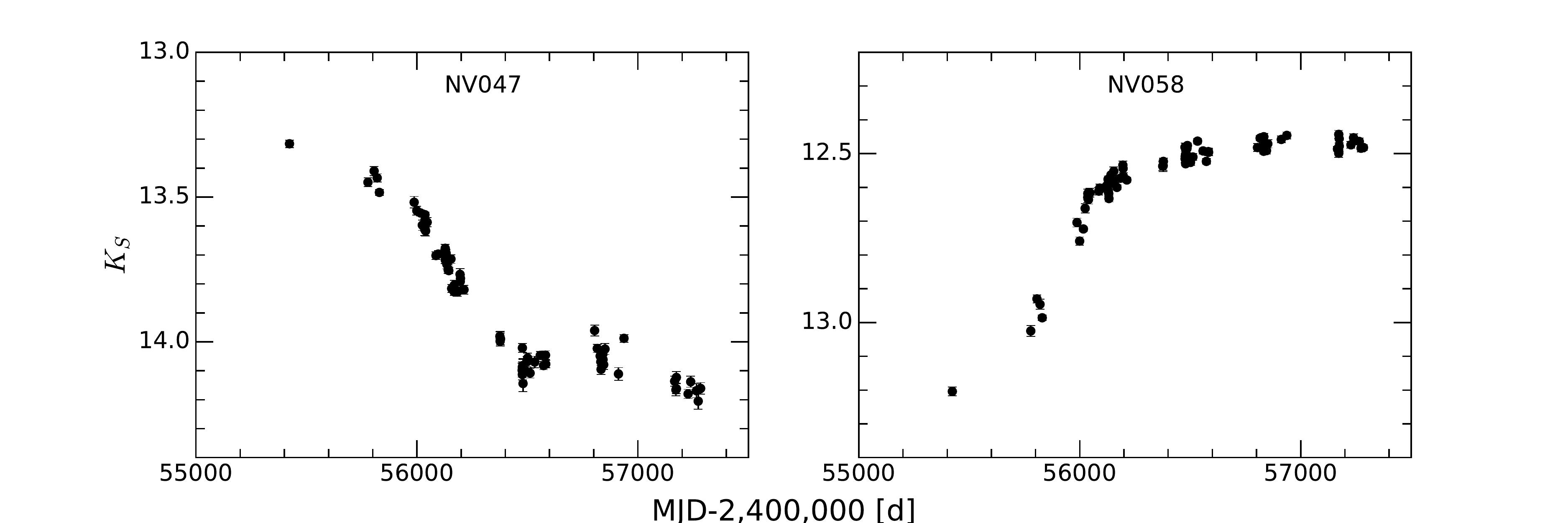}
\caption{Examples of red giant variables with long-term variations. (left) Light curve of NV047, a source that shows a decline of $\sim1$ mag in about 1000~days. (right) Light curve of NV058 displaying a steady brightness increase for all observed epochs.\label{fig-21}}
\end{figure*}

In Section~\ref{sec:lpv} we mentioned the presence of a group of stars with long-term variations that did not show light curves consistent with stellar pulsations. An example of these trends is illustrated in Figure~\ref{fig-21}. The left panel shows the light curve of NV047, a star with a strong dimming over the course of $\sim1000$~d. This trend is also observed in other stars, like NV043, NV044, NV062 and NV282. Sometimes, additional short-term fluctuations are observed along with the long-term decline, like in NV006 and NV193. In all these cases, the star goes down its brightness by about 1~mag in a timescale of $1000-2000$~d. 

The opposite effect is shown in the right panel of Figure~\ref{fig-21} for NV058. This variable manifests a continuous increase in brightness that lasts for the entire cycle of observations. A similar trend is present in other light curves, such as NV078, NV279 and NV325. 

Finally, other variables show a combination of dimming and brightening events, being NV228 the most extreme case. The GLS periodogram indicates a main period of 2080~d, along with a secondary period of 450~d. Nevertheless, this main period value is not reliable, since the star seems to keep its trend of rising brightness in the last epochs, so the real period could be even longer. 

The most plausible explanation for these long-term variations is related with extinction changes. For example, \citet{oliv01} observed a long-term periodic variation of 6180 d in the carbon-rich Mira V~Hya. The authors suggest that this variation was due to a dust cloud connected with a binary companion.

\begin{figure}
\centering
\includegraphics[width=\columnwidth]{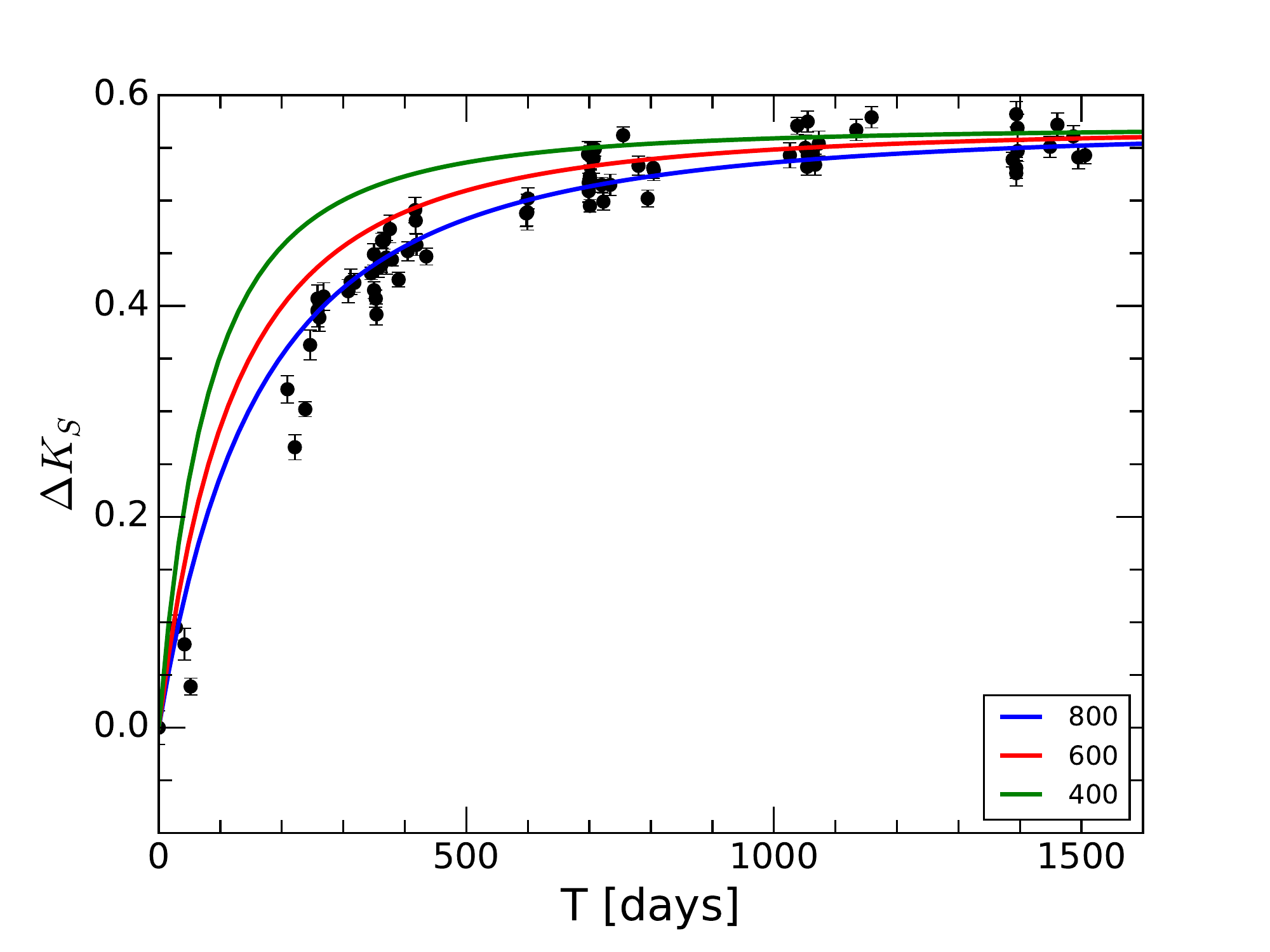}
\caption{Comparison of the brightening event of NV058 with the dust expansion model from \citet{whit97}. $T=0$ is the time when the obscuration begins to clear, and corresponds to the second epoch of observation. $\Delta K_S$ is the magnitude increase compared with the second epoch. Green, red and blue lines mark the models for initial radii of 400, 600 and 800~$\rsun$, respectively (see text for details).\label{fig-22}}
\end{figure}

Dust obscuration events are expected to occur in stars with thick circumstellar envelopes, and are caused by the strong mass loss that affects these objects, which continuously creates new dust \citep[e.g.,][]{whit97}. The brightening occurs when the obscuration ends. Models from \citet{wint94} predict that dust in C-rich Miras is expelled on every pulsational cycle and that it can cause long-term trends, as observed in the star R~For. Figure 11 from \citet{whit97} shows a portion of the light curve of this star, coupled with models of dust expansion at different initial radii. In this model, the dust moves at a speed of 20~km~s$^{-1}$, a value that the authors argue is representative for AGB stars.

We will replicate this model to match the light curve of NV058, a variable star that has a very similar behavior as the curve shown in Figure 11 of \citet{whit97}. The authors assume a uniform, spherically symmetric dust expansion with speed $v$ and a starting radius $R_0$. Therefore, the fractional change in magnitude $f$ with time $t$ will be

\begin{equation}
  f = 1 - (1-v\,t/R_0)^{-2} \,.
\end{equation}

The results of this simplified model are presented in Figure~\ref{fig-22}. The best-fitting model assumes an initial dust radius of 800 $\rsun$. This is further away than the starting radius of the dust for the variable R~For studied by \citet{whit97}, which is 500 $\rsun$. However, it seems to be more in line with the low temperature required to form carbon-rich dust above cool convective regions in RCB stars \citep{feas97b}.

A more realistic approach might be to assume that the dust is expelled in the form of random ``puffs''. If these puffs intersect the line of sight of the star, a dimming event is observed. This is similar to the obscuration phases that occur in RCB stars, though the timescales are different. A typical obscuration in RCBs lasts of order hundreds of days, while the comeback may take longer \citep[e.g.,][]{feas97a,mill12}. Variations observed in the majority of our long-term LPVs occur on a longer timescale.

If the reported trends are indeed caused by dust, we expect that colours might change along the reddening vector. However, this assumes that the extinction law for circumstellar material is the same as that for the interstellar medium, which is not completely valid, especially for C-rich stars \citep{ita11}. Hence the slope of the reddening vector for circumstellar extinction might be different. In any case, the colour changes observed in Figure~\ref{fig-8} are too small to be compatible with this idea.

\subsection{The spectrum of NV062}

\begin{figure*}
\centering
\includegraphics[width=\textwidth]{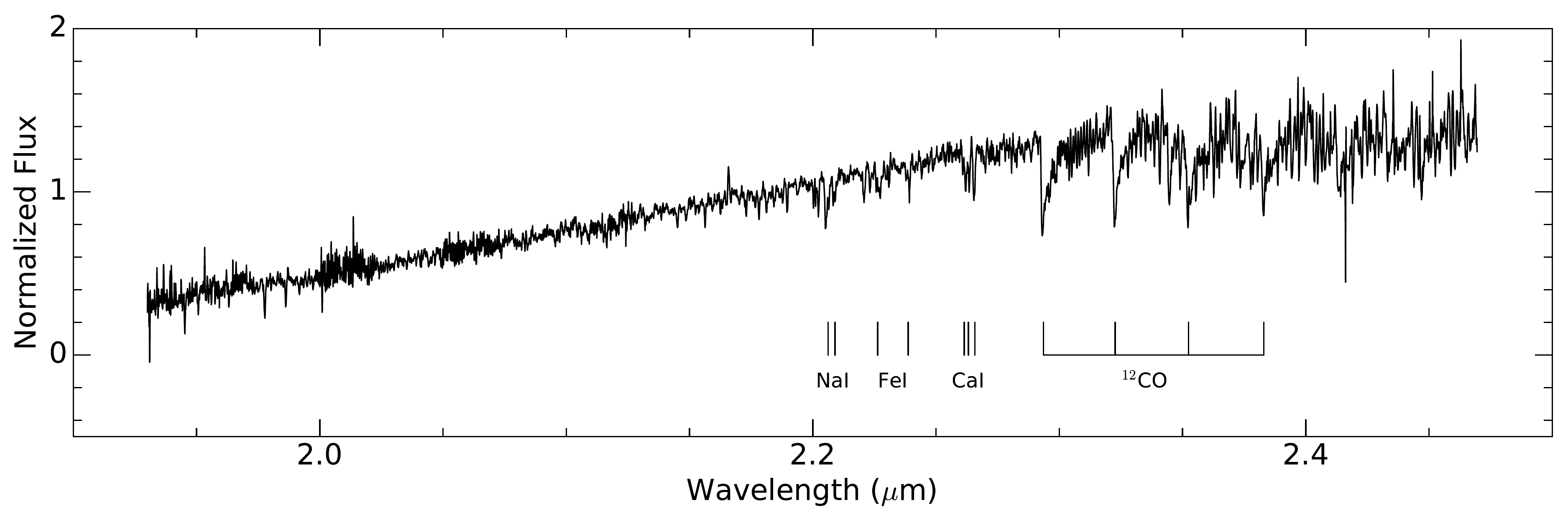}
\caption{$K_S$-band FIRE spectrum of NV062. Fluxes are normalized to the flux value corresponding to a wavelength of $2.2 \,\mu$m, and are not continuum corrected. The spectrum exhibits the typical CO, Na, Ca and Fe absorption features found in M-type stars.\label{fig-23}}
\end{figure*}

Object NV062 is one of the most interesting variables with long-term trends present in our catalog. The combined light curve shows a $\sim1$~mag increase during the \citetalias{mkn09} epochs, followed by a similar dimming during the epochs observed by the VVV survey. The complete time scale of this variation is $\sim5800$~d.

In order to get a better understanding of this variable star, it was observed in echellete mode with the FIRE spectrograph mounted on the Magellan Baade Telescope at Las Campanas Observatory, Chile, on 29.07.2017 (observer Ph. Lucas). The echellete mode provides coverage from $0.8-2.5\,\mu $m using echelle grating orders 11-32, at a resolving power of $R\sim 6000$ with a 0.6\arcsec\ slit. The spatial scale is 0.18\arcsec\,pixel$^{-1}$. Data reduction was performed using the echelle mode of the {\sc FIREHOSE} data reduction software. The final relative flux calibration and merging of individual orders was performed using the {\sc ONEDSPEC} tasks {\sc TELLURIC} and {\sc SCOMBINE} \citep[the complete procedure of data reduction is described in][]{cont17b}.

Figure~\ref{fig-23} illustrates the observed spectrum. The comparison with theoretical models taken from VOSA \citep{bayo08} gives the best fit results for AMES-Dusty \citep{alla01,chab00} and GRAMS-Orich \citep*{sarg11} models. A $T_{\rm eff}=3700\,$~K with a probability of 68\% is obtained. The GRAMS-Orich models also compute $L = 3.8\times10^4 \, L_{\odot}$ and $\log(g)=-0.5$. In addition, a metallicity of ${\rm [Fe/H]}=-0.44\pm0.18$~dex is obtained. The comparison with template spectra of stars with known spectral type reveals a class of M4-5\,III with a probability of 72\%. Thus, we confirm the nature of a typical O-rich AGB star with dust shells around.

The previous result might seem surprising, since we expect that C-rich LPVs show long-term variations caused by dust obscuration. This is due to the large opacity of carbon grains \citep{laga08}. However, long-term trends are also observed~-- though to a lesser extent~-- in O-rich LPVs \citep{whit06}. Since only a handful of C-rich Miras have been found in the Galactic bulge so far \citep{mats17}, it is safe to assume that the majority of these LPVs are O-rich. Hence, it seems that obscuration events in O-rich stars may be more common than previously realized. 

Another way to look at this is the possibility that~-- at the distance of the GC~-- the VVV photometry is more sensitive to red giants with thick circumstellar envelopes. AGB variables with thinner dust shells will probably be saturated in our images, as discussed previously. This is supported by the magnitude distribution of our Mira variables, and the dereddened colours of red giants shown in Figure~\ref{fig-16}, where the majority of these stars have $(J-K_S)_0>2$~mag.

In conclusion, since the timescales for these events can be several years, it is necessary to observe them over a longer temporal baseline, in order to fully characterize these long-term trends, and get a better understanding of their variability mechanisms.


\subsection{Analyzing 14 years of observations}

\begin{figure*}
\begin{center}
\begin{minipage}{0.5\textwidth}
  \centering
  \includegraphics[width=\textwidth]{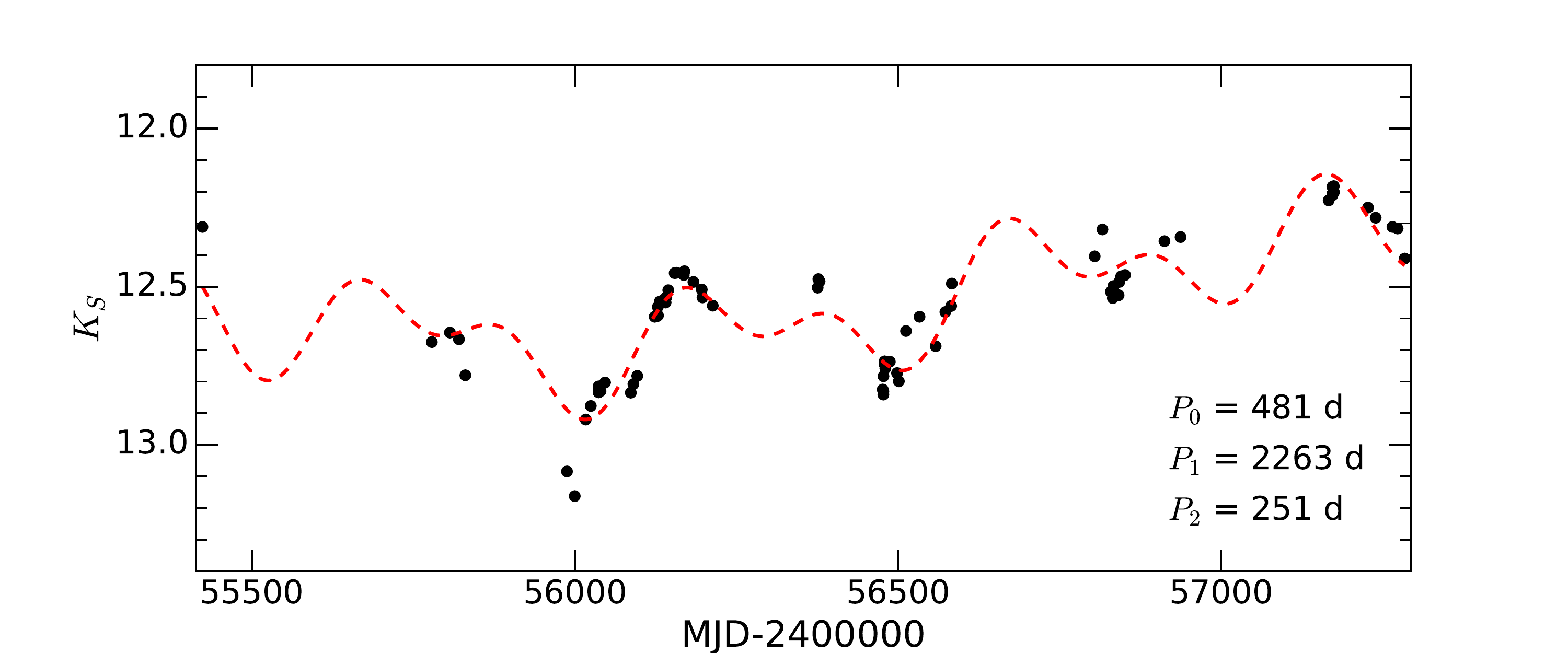}
\end{minipage}\hfill
\begin{minipage}{0.5\textwidth}
  \centering
  \includegraphics[width=\textwidth]{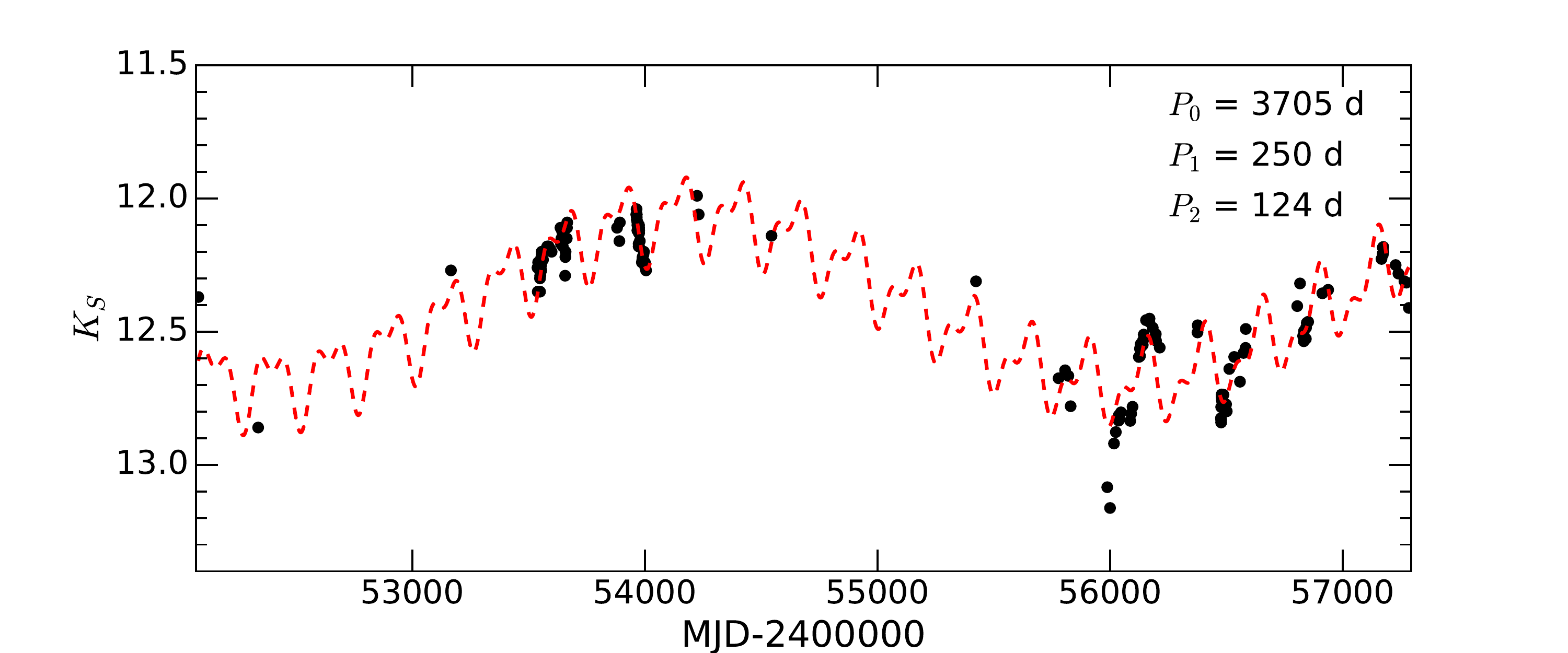}
\end{minipage}\hfill
\caption{(top) Light curve for the SR variable NV096. Multiple periods were obtained using {\sc Period04}. The corresponding Fourier fit is marked as a red dashed line. (bottom) Same calculation, but for the combined data. The first three periods fitted are shown in each panel.\label{fig-24}}
\end{center}
\end{figure*}

The combination of the light curves from \citetalias{mkn09}, \citetalias{mfk13}, and the present work makes it possible to study the long-term variations of these stars. Observations go from 2001 to 2015, with the majority of epochs concentrated between 2005-2006 and 2012-2015. Between 2008 and 2010 only two observations were carried out, the first one with IRSF and the second with VISTA. There were no observations during 2009. Nevertheless, the combined light curves have between 118 and 173 epochs in the $K_S$-band, a sufficient number to perform a thorough analysis of their variations. We must note that a shift of up to $\pm0.2$~mag might be present between both data sets (see Section~\ref{sec:literature}). However, the main conclusions of the following analysis should not be affected by these differences.

There are 41 variables that do not have regular periodicity. Light curves show a wide variety of shapes. There are some stars that have larger amplitudes in one dataset compared to another, such as NV063, NV180, and NV311. Other variables, like NV096 and NV311, have light curves that go from semi-regular to irregular variations, or vice-versa. This could be explained by a pulsation mode shift \citep{kiss00,tem05}. Another possible explanation is given by \citet{harti14}. Through the use of high-precision {\it Kepler} photometry, the authors found that SR variables pulsate in multiple modes, mainly the first overtone coupled with the fundamental one. In addition, they reported that the given period assigned to an SR star could change depending on the time interval selected to measure that period, and that some stars that were historically classified as IRR showed clear periodicity in the {\it Kepler} data set. \citeauthor{harti14} suggest that these irregular phases might be caused by the superposition of multiple periods within a narrow range of values. Depending on the combination of these periods, the resulting light curve could show up as irregular or (quasi-)periodic. A similar situation is also observed in OSARGS \citep{sosz04}.

The already mentioned source, NV096, is classified as an SR variable. In fact, multiple periods are observed in this star, as depicted in Figure~\ref{fig-24}. Using only the VVV data, this star shows a main period of 481~d and a long secondary period of 2189~d. However, if we add the observations from \citetalias{mkn09}, the periods derived change significantly, and a main period of 3705~d arises. An interesting fact is that a periodic signal of 250~d is in common between both sets of observations. 

Another type of variability found through the inspection of the combined light curves corresponds to long-term brightening and/or dimming episodes, such as the already discussed cases of NV058 and NV062. The typical magnitude change is about 1~mag. There are some cases where this variation is more extreme. NV232 has a steady brightening of $\sim1$~mag during the \citetalias{mkn09} epochs, followed by a 5-year fading of $\sim2.5$~mag during our observations. NV261 undergoes a similar dimming. As suggested in the beginning of this Section, the most probable causes behind these variations are dust obscuration events. 

Regular variables show, in general, a good agreement in their combined light curves. The infrared Miras, namely NV106, NV118, NV188, and NV295, exhibit amplitude and, in some cases, period variations across their cycles. For NV106, the period that we calculated using the \citetalias{mkn09} data is 445~d, compared with 437~d as obtained from our observations. Similarly, the period for NV118, estimated only with \citetalias{mkn09} data, is 860~d, though in this case the period is poorly constrained, since there is missing information about its maximum brightness, and only one pulsation cycle is observed. We must note that the \citetalias{mkn09} and \citetalias{mfk13} observations are mostly concentrated between 2005 and 2006, hence their research cannot properly characterize Miras with periods longer than $\sim700$~d.

Even with the larger time frame of the combined data sets, the total number of observed cycles for these Miras is small. For example, NV118 would complete four cycles in these 14 years. In consequence, we cannot conclude that these period variations are part of longer-term secular changes as those studied by \citet{tem05}. Nevertheless, the period changes and the timescales are consistent with the already described meandering Miras \citep{zij02}.

On the other hand, the light curve of NV188 presents noticeable cycle-to-cycle amplitude variations that are suggestive of a C-rich Mira \citep{whit03,whit06}, a situation also observed in NV315. However, neither of these variables have enough photometric information in other IR bandpasses to check if their colours are compatible with C-rich photospheres.

As pointed out by the OGLE team  (see for example Soszynski et al. 2014), most of the LPVs are multi-periodic (up to three periods are published in the OGLE-III catalogues). The same trend is found in our sample. 

The remainder of the regular variables show a very good agreement between both data sets.


\section{YSO candidates}

\begin{figure*}
\centering
\includegraphics[width=\textwidth]{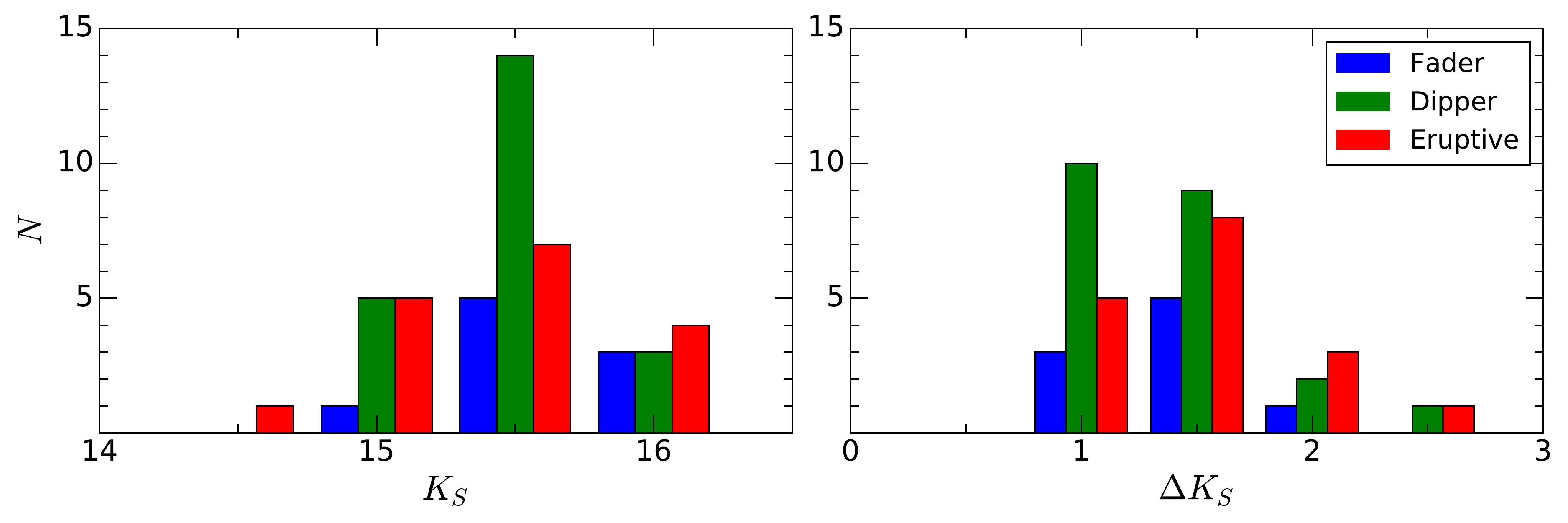}
\caption{(left) Histogram for YSO candidates found in the present work. (right) Distribution of amplitudes for YSO candidates. In both panels, blue, green and red bars correspond to faders, dippers and objects with eruptive variability, as defined in \citet{cont17a}. \label{fig-25}}
\end{figure*}

The majority of the remaining variable stars present in this catalog are classified as YSO candidates. With the available photometric information, we cannot be certain about their classification. These are very faint stars, as expected for a YSO at the distance of the GC, so for most candidates there are no magnitudes other than in the $K_S$-band. Hence, they are classified solely based on their light curve morphology.

From a purely morphological standpoint, the distinction between YSOs related with extinction events~-- faders and dippers~-- and LPVs with long-term variations is not simple. Indeed, we observe similarities between a fader and the light curves of red giants undergoing dust obscuration events. A first distinction is carried out through their colours and magnitudes. However, except for some $H$ magnitudes, we do not have colour information for these variables, and the $H-K_S$ colour by itself is not sufficient to clearly distinguish between both types. In order to establish a mean magnitude cut, we analyzed the pre-main sequence isochrones discussed in \citetalias{nav16}. A reasonable lower magnitude for YSOs located at the distance of the GC is $K_S\approx15$~mag. However, we cannot rule out the possibility of finding YSOs closer than the GC, whose magnitudes will be brighter. In this regard, we expect some confusion in the classification of these sources with similar light curves.

There are 48 YSO candidates in our sample, 22 classified as dippers, 9 as faders and 17 with eruptive variability, according to the criteria explained in \citet{cont17a}. The magnitude and amplitude distributions for these stars are shown in Figure~\ref{fig-25}. The mean magnitude of these YSOs has a peak at $K_S=15.5$~mag, which is consistent with them belonging to the GC. Regarding their amplitudes, most of these sources have $\Delta K_S$ between 1 and 1.5~mag. 

Among the YSO candidates with eruptive variability, we highlight the cases of NV035, NV125, NV141, and NV185. All their light curves show a strong brightening of $\sim1.5$~mag that is consistent with FUor-like variables, such as those reported by \citet{cont17a}. On the other hand, NV037 and NV345 show episodic burst events.

A spectroscopic follow-up will be required in order to obtain a certain classification for these candidates.


\section{Summary}

We reported a new catalog with the variable stars population located near the GC. The main difference between this work and previous studies is that we were able to reach fainter magnitudes, down to $K_S=16.5$~mag, allowing the analysis of a more varied population, including dust-enshrouded AGB stars and bright YSO candidates, among others. We remind the reader that our data do not allow an investigation of the brightest stars in the field.

We detected 353 variable star candidates in a region of $11.5'\times11.5'$ that comprises two of the most massive young star clusters in our Galaxy: Arches and Quintuplet. This is a complement of the 33 sources analyzed within 2 arcmin from the centre of Quintuplet, that were discussed in \citetalias{nav16}. Eighty-five percent of these stars (299 objects) have not been reported in previous variability studies carried out in the GC. 

The large majority of the variable sources detected are red giant stars, most of them belonging to the AGB phase. Aside from the typical Miras, SR and IRR variables widely described in the literature, we found a sizable sample of giant stars with long-term trends. We have analyzed the probable causes of these variations.

Among the AGB stars present in our catalog and in \citetalias{nav16}, we have 13 Mira variables. The majority of them have periods longer than 400~d, implying that they are infrared Miras, as defined in this work. We derived periods and extinctions for 9 of these sources. When the \citetalias{nish06} or \citetalias{alon17} extinction laws are used, Miras with periods shorter than 400~d, for which the PL relation is well defined, have distances placing them close to the GC. Longer-period Miras have less constrained distances, due to the strong circumstellar extinction that affects them. This is an indication of very large mass loss rates \citep{wood07}. One of these Miras seems to be located far beyond the Galactic bulge.

We were able to use near- and mid-IR colours, combined with dust radiative transfer models, in order to infer the atmospheric C/O ratio of AGB stars. For a small subset of stars we could classify them as probable C-rich or O-rich stars. O-rich symbiotic stars also show very long period variations, as Carbon stars do (\cite{2013MNRAS.432.3186M}). 
Other AGB stars have strong IR excesses coming from their dust shells that make it hard to distinguish between both chemistries, without access to longer-wavelength magnitudes and/or spectroscopy.

We don't have any variables in common with Mowlavi et al. (2018) catalog of LPVs based on Gaia DR2, thus we consider the catalog presented here as an extension towards the crowded and obscured Galaxy regions.

Finally, we detected 48 objects classified as YSO candidates, according to their light curve morphology. This is the first time that a search for YSOs is carried out in this field. 


\section*{Aknowledgements}

C.N.-M., J.B., M.C., N.M, R.C, and D.M. gratefully acknowledge support by the Ministry for the Economy, Development, and Tourism's Millennium Science Initiative through grant IC\,120009, awarded to the Millennium Institute of Astrophysics (MAS). D.M. and M.C. acknowledge support from CONICYT project Basal AFB-170002. D.M. is also supported by FONDECYT grant \#1170121.  M.C. additionally acknowledges support by CONICYT/RCUK's PCI program through grant DPI20140066; by FONDECYT grant \#1171273; and by the DAAD and DFG (Germany). This research is based on observations made with ESO Telescopes at the Paranal Observatory under programme ID 179.B-2002. We thank anonymous referee for useful comments and suggestions. PWL acknowledges support by the UK Science and Technology Facilities Council under consolidated grants ST/R000905/1 and ST/M001008/1. CCP acknowledges support from the Leverhulme Trust.






\bsp	
\label{lastpage}
\end{document}